\documentclass[twocolumn,prl,reprint]{revtex4-2}
\usepackage[english]{babel}
\usepackage{verbatim}
\usepackage{float}
\usepackage{amsmath}
\usepackage{amsthm}
\usepackage{amssymb}
\usepackage{graphicx}
\usepackage{wasysym}
\usepackage{esint}
\usepackage{wasysym}
\usepackage{xcolor}

\makeatletter

\usepackage{amsmath}

\renewcommand{\theequation}{\arabic{equation}}

\makeatother

\begin{document}
\title{
Odd-viscosity-induced instability in shear flows
}

\author{Yonatan Messica}
\affiliation{
	Department of Physics, Bar-Ilan University, Ramat Gan, 52900, Israel
}
\author{Igor Gornyi}
\affiliation{\mbox{Institute for Quantum Materials and Technologies, Karlsruhe Institute of Technology, 76021 Karlsruhe, Germany}}
\affiliation{
Institut f\"ur Theorie der Kondensierten Materie, Karlsruhe Institute of Technology, 76128 Karlsruhe, Germany}
\author{Dmitri B. Gutman}
\affiliation{
	Department of Physics, Bar-Ilan University, Ramat Gan, 52900, Israel}

\begin{abstract}
Odd viscosity is a nondissipative component of the viscosity tensor that arises in fluids with broken time-reversal symmetry.
We show that, despite being energy-conserving, odd viscosity can qualitatively alter hydrodynamic stability by generating exponentially growing modes that are absent in conventional fluids. 
For plane Poiseuille flow, we derive the odd-viscous generalization of the Orr--Sommerfeld--Squire equations and find a new instability that first emerges for streamwise-uniform modes and extends to oblique modes. 
The instability exploits the non-normality of the linearized Navier-Stokes operator: conventional fluids support transiently growing disturbances through the lift-up mechanism,
while odd viscosity closes a feedback loop between wall-normal velocity and vorticity, converting these disturbances into self-sustaining exponentially growing modes.
This mechanism lowers the linear instability threshold by nearly two orders of magnitude, from $\mathrm{Re}_c \approx 5772$ to $\mathrm{Re}_c \approx 62$.
More generally, we show that an energy-conserving coupling can destabilize a linearly stable
system by sustaining the growth of otherwise transiently growing disturbances.
Such a coupling thus triggers energy growth without directly supplying energy,
by enabling continuous energy extraction from the background state.
This establishes a mechanism for controlling stability in non-normal systems without direct energy dissipation or injection.

\end{abstract}
\maketitle

\section{Introduction}
\label{Introduction}
\textbf{}%
\textbf{}%

Odd viscosity is a nondissipative linear response of the stress tensor to velocity gradients \cite{avron_odd_1998, fruchart_odd_2023}.
Historically, it was first recognized in plasma physics, where it is known as gyroviscosity \cite{rosenbluth1962nuclear,roberts_magnetohydrodynamic_1962}. 
There, it arises from finite-Larmor-radius effects and plays an important role in the dynamics and stability of magnetically confined and space plasmas.
In electronic fluids, odd (Hall) viscosity belongs to the broader family of nondissipative Hall responses associated with broken time-reversal symmetry.
Much like the intrinsic anomalous Hall conductivity, it has a geometric origin and
is intimately connected to the Berry curvature of the underlying electronic states \cite{avron_viscosity_1995,read_non-abelian_2009,srivastava_electron_2025}.
 
From the symmetry perspective, odd viscosity requires breaking of time-reversal symmetry, as dictated by the Onsager reciprocity relations. 
This requirement can be realized in a variety of physical systems, including active fluids composed of self-spinning particles 
\citep{banerjee_odd_2017,soni_odd_2019,markovich_odd_2021} and electronic fluids, 
where time-reversal symmetry is broken either by an external magnetic field \citep{berdyugin_measuring_2019} 
or spontaneously in anomalous Hall systems \citep{li_electronic_2026}. 
Despite its appearance in such diverse physical systems, odd viscosity has not become part of the standard framework of classic hydrodynamics. 
In contrast to ordinary viscosity, whose influence on fluid flows is treated systematically in hydrodynamic textbooks, 
the role of antisymmetric, nondissipative viscosity remains much less understood. 
This raises a fundamental question: what general statements can be made about the influence of odd viscosity on fluid motion? 
 
One of the most fundamental problems in fluid mechanics is hydrodynamic
stability, which arises across all length scales, from microfluidics
to astrophysical flows \citep{drazin_hydrodynamic_2004}. Odd viscosity
has already been shown to have a profound influence on hydrodynamic
instabilities in various systems. Depending on the physical setting, it may either stabilize
or destabilize the flow, as demonstrated for thin films
\citep{kirkinis_odd-viscosity-induced_2019,mukhopadhyay_thermocapillary_2021,zhao_effect_2021,samanta_role_2022,hossain_stability_2025,tong_influence_2026},
magnetized plasmas
\citep{rosenbluth1962nuclear,steinhauer_gyroviscous_1990},
and Taylor--Couette flow \citep{du_stability_2023}. In turbulent
flows, odd viscosity modifies the energy cascade through the emergence
of odd waves \citep{de_wit_pattern_2024,chen_odd_2024,de_wit_wave_2026}. Despite these developments,
its influence on one of the canonical problems of hydrodynamic
stability -- the stability of incompressible, isothermal parallel
shear flows -- remains unexplored.

The role of ordinary, dissipative viscosity in the stability of parallel shear flows is surprisingly rich. 
At low Reynolds numbers, viscosity suppresses perturbations and stabilizes laminar flow. 
Yet, viscosity can also be the very origin of instability. A classic example is the Tollmien--Schlichting instability: 
whereas shear flows without an inflection point are linearly stable in the inviscid limit, 
they may become unstable once an arbitrarily small viscosity is introduced \citep{drazin_hydrodynamic_2004}. 
The understanding of this phenomenon has been indispensable for modern aerodynamics.

Given the profound and multifaceted role of dissipative (even) viscosity in hydrodynamic stability, 
it is natural to ask about its odd counterpart -- does odd viscosity
merely modify the quantitative characteristics of classic instabilities,
or does it give rise to qualitatively new instability mechanisms?
To answer this question, we investigate the stability of plane
Poiseuille flow in an incompressible, odd viscous fluid.

Because odd viscosity conserves energy, one might naively expect it to be irrelevant for hydrodynamic stability, 
which is associated with the growth of perturbation energy. Here we show that this intuition is fundamentally wrong:
despite conserving energy, odd viscosity destabilizes the flow over a large region of the perturbation parameter space.
Crucially, the instability arises even for odd viscosity that is smaller in magnitude than dissipative viscosity,
as is the case for experimental realizations of odd viscosity to date \cite{soni_odd_2019,berdyugin_measuring_2019}.
The destabilizing effect is due to odd viscosity converting transiently growing
disturbances into self-sustaining, exponentially growing modes.
This remarkable physical mechanism is discussed for
general energy-conserving perturbations
in Section \ref{mechanism} of the paper.

\section{Results}

\subsection{Odd Orr--Sommerfeld--Squire equation}
\label{Odd Orr--Sommerfeld--Squire }
We begin by extending the standard theory of linear stability analysis, namely the Orr--Sommerfeld--Squire equations \cite{schmid_stability_2001}, to fluids with odd viscosity.
In general, the viscosity tensor of a fluid describes its stress response to velocity gradients,

\begin{equation}
\sigma_{ij}=\eta_{ijkl}\partial_l{v}_{k},
\end{equation}
where $\sigma$ is the viscous stress tensor, $\boldsymbol{v}$ is the velocity field,
$\partial_i\equiv\partial/\partial x_i$ is a spatial derivative,
and $\eta$ is the viscosity tensor. 
Odd viscosity refers to the part of the viscosity tensor that
is antisymmetric under the exchange of the index pairs, $\eta_{ijkl}^{{\rm o}}=-\eta_{klij}^{{\rm o}}$ \citep{avron_odd_1998}.
Throughout this work, we consider the odd viscosity tensor
of the form 
\begin{equation}
\eta_{ijkl}^{{\rm o}}=\frac{\eta^{{\rm o}}}{2}\left(\delta_{ik}\epsilon_{jlz}+\delta_{jl}\epsilon_{ikz}+\delta_{jk}\epsilon_{ilz}+\delta_{il}\epsilon_{jkz}\right),
\end{equation}
where the distinguished axial direction is along the $z-$axis, either due to an intrinsic rotation of the fluid constituents \citep{markovich_odd_2021} 
or an external magnetic field \citep{reynolds_three_2023}.
In addition to the odd viscosity contribution, the viscosity tensor contains the conventional shear viscosity term  $\eta^{\rm e}$.


We study the linear stability of a three-dimensional incompressible shear flow with base velocity
$\boldsymbol{V}=V(y)\hat{\boldsymbol{x}}$. For the odd viscosity tensor introduced above, whose distinguished axis is the $z$ axis, odd viscosity does not modify the base-flow profile; its only effect is a trivial renormalization of the pressure. 
We consider normal-mode perturbations of the form
\begin{equation}
\boldsymbol{u}=\boldsymbol{u}(y)e^{i \left( \alpha x+\beta z - \omega t \right) },
\end{equation}
and linearize the Navier--Stokes equations about the base flow. 
The resulting generalized Orr--Sommerfeld--Squire problem
(derived in Appendix \ref{appendix oss-derivation})
is formulated in terms of the wall-normal velocity $u_y$ and the wall-normal vorticity $\varpi_y$,
and is given by

\begin{equation}
\begin{aligned}
&L\begin{pmatrix}
u_y\\
\varpi_y
\end{pmatrix}
=
i \omega
\begin{pmatrix}
- \Delta_k & 0\\
0 & 1
\end{pmatrix}
\begin{pmatrix}
u_y\\
\varpi_y
\end{pmatrix},
\\[2mm]
&L \equiv
\begin{pmatrix}
L_{\rm OS} &
-i\beta\frac{1}{\rm Ro}\Delta_k\\
i\beta\left(V'
-\frac{1}{\rm Ro}\Delta_k\right) &
L_{\rm SQ}
\end{pmatrix}.
\end{aligned}
\label{eq:OS-SQ odd matrix operator}
\end{equation}

Here, $k=\sqrt{\alpha^2+\beta^2}$ is the wavevector magnitude and
$\Delta_k=\partial_y^2-k^2$ is the partially Fourier-transformed Laplacian.
We further introduce the conventional Reynolds number ${\rm Re}$ and the odd Reynolds number ${\rm Ro}$, 
characterizing the ratios of the inertial forces to the shear and odd viscous forces, respectively,
\begin{align}
\rm Re &= \bar{V}\delta/\nu^{{\rm e}}, \\
\rm Ro &= 2 \bar{V}\delta/\nu^{{\rm o}},
\end{align}
where $\bar{V}$ is the velocity scale, $\delta$ is the channel half-width,
and $\nu^{\rm{e(o)}}\equiv\eta^{\rm{e(o)}}/\rho$
are the kinematic viscosities, $\rho$ being the fluid density.
The Orr--Sommerfeld and Squire operators are
\begin{align}
L_{{\rm SQ}} & =i\alpha V-\frac{1}{{\rm R^{{\rm e}}}}\Delta_{k},\\
L_{{\rm OS}} & =-L_{{\rm SQ}}\Delta_{k}+i\alpha V^{\prime\prime}.
\end{align}

In contrast to the conventional Orr--Sommerfeld--Squire formulation, 
odd viscosity couples wall-normal velocity and vorticity in both directions,
thereby modifying both the eigenvalue spectrum and the stability properties of the flow.

Supplementing Eq. (\ref{eq:OS-SQ odd matrix operator}) with boundary conditions results in a
generalized eigenvalue problem.
For channel-bounded flows with no-slip, impermeable walls at $y=\pm1$, the
boundary conditions are given by
\begin{equation}
u_y(\pm1)=
\left.\frac{\partial u_y}{\partial y}\right|_{y=\pm1}
=\varpi_y(\pm1)=0.
\end{equation}

Equation (\ref{eq:OS-SQ odd matrix operator})
constitutes the generalization of the conventional Orr--Sommerfeld--Squire equations to three-dimensional flows with odd viscosity.
For two-dimensional perturbations ($\beta=0$),
the odd viscosity contribution vanishes,  in agreement with
the known result of Ref. \citep{ganeshan_odd_2017}. In the absence
of odd viscosity ($1/{\rm Ro}=0$), the Orr--Sommerfeld--Squire system
is well studied \citep{schmid_stability_2001}, 
and a brief review of its properties
is presented in Appendix \ref{appendix classic OSS eqs}. An important feature
of the standard case is the triangular structure of  the operator,
allowing separation of the eigenmodes into two families of modes,
Orr-Sommerfeld and Squire modes.
Finite odd viscosity destroys this triangular structure, so the eigenmodes
can no longer be classified purely as Orr-Sommerfeld or Squire modes.

\subsection{Linear stability analysis of plane Poiseuille flow}
\label{Linear stability analysis}
We now apply Eq. (\ref{eq:OS-SQ odd matrix operator})
to the canonical problem of plane Poiseuille flow,
\begin{equation}
V(y)=1-y^2,\qquad -1\le y\le1.
\end{equation}
We solve the generalized Orr--Sommerfeld--Squire eigenvalue problem 
numerically for different values of the parameters $\left({\rm Re},{\rm Ro},\alpha,\beta\right)$.

To determine the stability of the flow, we focus on the eigenvalue with the largest imaginary part.
The parameter values for which this eigenvalue becomes purely real define the marginal-stability curves, separating stable flows,
in which all perturbations are damped, from unstable flows, in which at least one perturbation grows exponentially.

The Poiseuille problem is invariant under reversal of the sign of odd viscosity;
the eigenvalue spectrum is unchanged, while the corresponding eigenmodes are
related by a symmetry transformation (see Appendix \ref{appendix symmetry transformation}). 
Therefore, we present results only for positive values of ${\rm Ro}$.
\subsection{Odd-viscosity-induced instability}

\begin{figure}[H]
\begin{centering}
\includegraphics[scale=0.35]{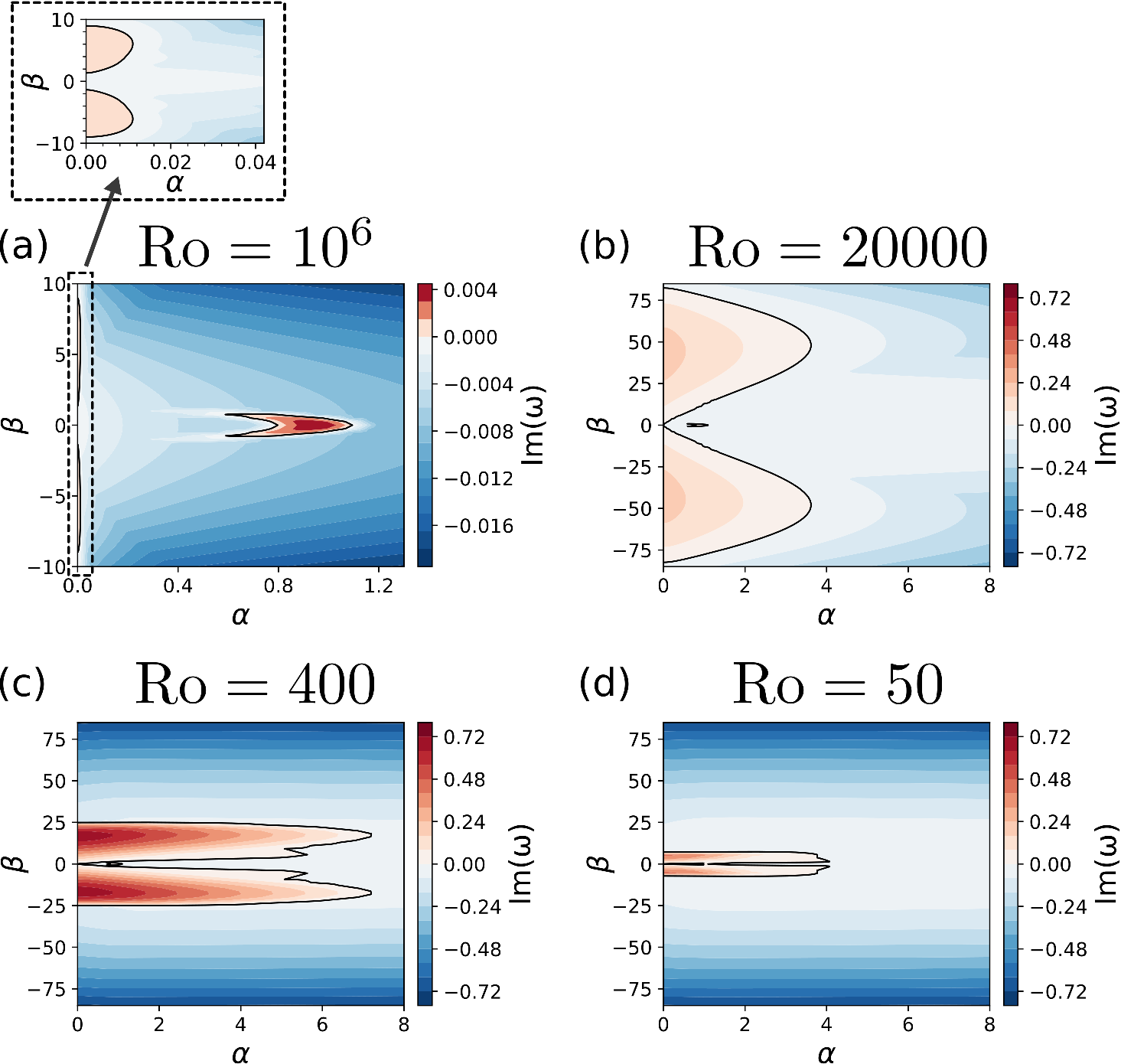}
\par\end{centering}
\caption{Growth rate of the most unstable mode [largest $\rm{Im}\left( \omega \right) $ value] in the $\alpha-\beta$
plane for ${\rm Re=10000}$, (a) ${\rm Ro=10^6}$, (b) ${\rm Ro=20000}$,
(c) ${\rm Ro=400}$, and (d) ${\rm Ro=50}$. The black outlines mark the curves of marginal stability $\rm{Im}\left( \omega \right) =0$.
Inset of (a) shows a zoomed view near the line $
\alpha=0$, where the odd viscous instability emerges.
\label{fig:growth rate alpha-beta plane}}
\end{figure}

Fig. \ref{fig:growth rate alpha-beta plane}
shows  the linear stability diagram in the
$\left(\alpha,\beta\right)$ parameter plane for ${\rm Re=10000}$
and different values of ${\rm Ro}$. For the following, it is useful to
consider the streamwise phase velocity corresponding to the eigenfrequency of a mode,
\begin{equation}
c=\frac{\omega}{\alpha}.
\end{equation}

In the absence of odd viscosity, there is a single unstable
region of streamwise-propagating waves, corresponding
to the classic Tollmien-Schlichting instability. 
The unstable mode is localized near the channel walls and propagates with phase velocity 
$c_{r}\approx0.26$ (writing the complex phase velocity as $c=c_{r}+i c_{i}$). It is characterized by a critical layer at $y=y_{c}$, 
where the local base-flow velocity equals the wave phase velocity
 $V(y_{c})=c_{r}$.


As the value of odd viscosity exceeds a critical threshold, a new unstable
region appears. The first modes to become unstable are the ones modulated in the
spanwise direction (perpendicular to both the streamwise and wall-normal
directions), corresponding to the vicinity of the line $\alpha=0$ in
Fig.~\ref{fig:growth rate alpha-beta plane}. As odd viscosity
increases further, the unstable region expands toward finite values
of $\alpha$, indicating that unstable wavevectors become progressively
more aligned with the streamwise direction. 


For intermediate odd viscosity
($1/{\rm Ro}\gtrsim1/{\rm Re}$), the unstable region contracts along
the spanwise direction while continuing to expand toward the
streamwise direction. At the same time, the maximum growth rate of the
unstable modes increases. For sufficiently strong odd viscosity, this
trend reverses and odd viscosity acts purely as a stabilizing
mechanism. Eventually, at sufficiently large odd viscosity, the flow becomes
linearly stable with respect to all modal perturbations [except
for a vanishing area in the phase space near
the Tollmien-Schlichting instability of streamwise modes ($\beta=0$),
where the odd viscous force vanishes].

We next examine the full eigenvalue spectrum of 
the odd Orr--Sommerfeld--Squire (OSS) system.
In Fig. \ref{fig:eigval plot}, we plot the eigenvalues
for varying values of ${\rm Ro}$, for (a) oblique propagating modes ($\alpha=\beta$)
and (b) near-spanwise modes ($\beta \approx k$).
In the absence of odd viscosity, the spectrum of the Orr-Sommerfeld-Squire operator
contains many groups of nearly degenerate eigenvalues,
corresponding to damped Orr-Sommerfeld and Squire modes
\cite{chapman_subcritical_2002}.
Odd viscosity hybridizes the modes and lifts the (almost) degeneracy.
For the oblique modes in Fig. \ref{fig:eigval plot}(a), the odd-viscosity-induced instability  
emerges for a mode with phase velocity $c_r\approx0.92$,
and for the near-spanwise modes in  Fig. \ref{fig:eigval plot}(b), the instability arises
for a mode with velocity $c_r\approx0.58$. These correspond to modes
with a critical layer localized in the interior of the channel,
in contrast to the Tollmien--Schlichting modes,
whose critical layer lies close to the
channel walls.

It is interesting to identify the origin of an unstable mode
with the mode that adiabatically evolves into it from
another point in the parameter space.
As can be seen in Fig. \ref{fig:eigval plot},
the spectrum evolves continuously when a single parameter is changed
(in the plot, $\rm{Ro}$), uniquely defining a one-to-one mapping
between the eigenmodes at different values of $\rm{Ro}$.
However, the evolution of the spectrum becomes path-dependent
when varying more than one parameter in the OSS equation;
that is, the adiabatic evolution mapping between the two sets of eigenvalues corresponding to
two parameter pairs [e.g., $\left( \rm{Ro}_1, \alpha_1 \right)$ and $\left( \rm{Ro}_2, \alpha_2 \right)$]
is not uniquely defined by the endpoints.
This is a signature of exceptional points that arise for the odd OSS equation at specific
parameter values \cite{berry_physics_2004,heiss_exceptional_2004,kern_subharmonic_2022},
another hallmark of its non-Hermitian nature. We elaborate on this and demonstrate
the path dependence of the spectrum in Appendix \ref{Exceptional pts appendix}.
It is worth mentioning a recent work \cite{kirkinis_mathscrpt-symmetric_2026}
studying the consequences of parity-time (PT) symmetry in odd viscous fluids.
There, a driven odd viscous system is presented that
exhibits exceptional points which separate PT-unbroken and PT-broken phases.

\begin{center}
\begin{figure}[H]
\begin{centering}
\includegraphics[scale=0.28]{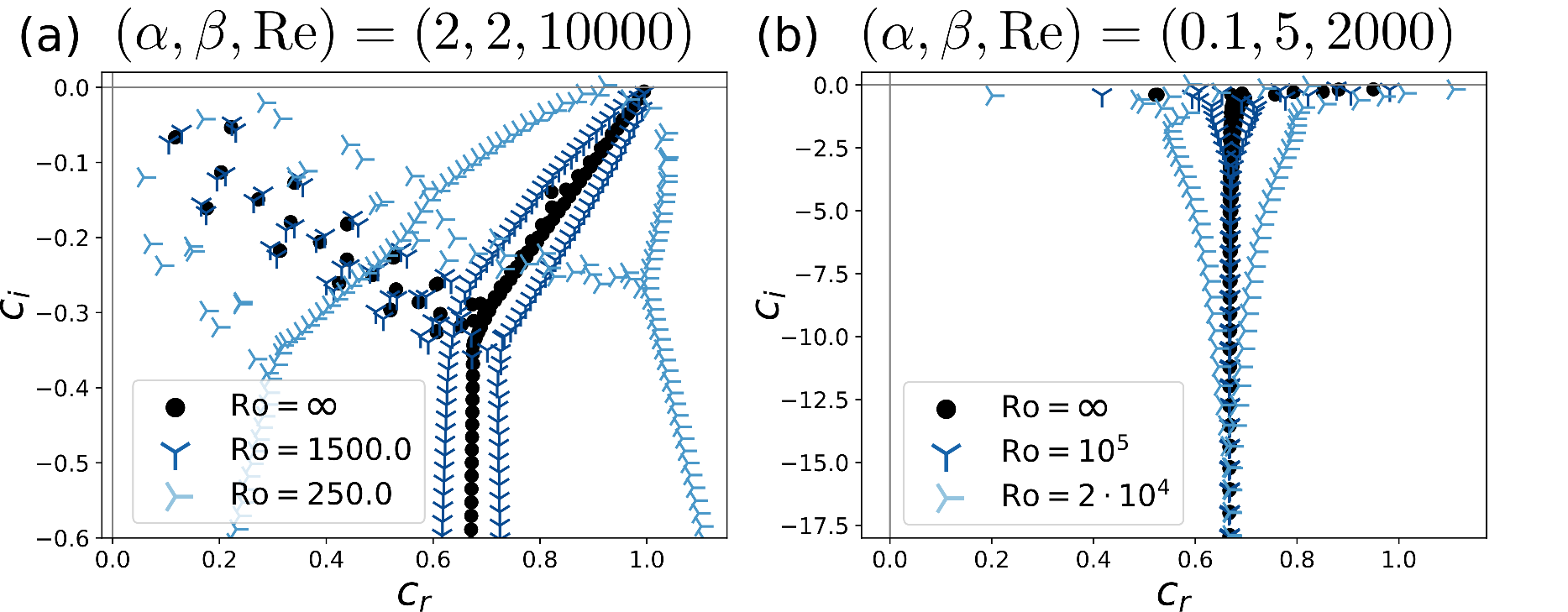}
\par\end{centering}
\caption{
Eigenvalues $c=c_{r}+i c_{i}$ for varying values of $\rm{Ro}$ for (a) oblique modes ($\alpha=\beta=2, \rm{Re}=10000$), and (b) near-spanwise modes ($\alpha=0.1,\beta=5, \rm{Re}=2000$).
In the absence of odd viscosity, the spectrum has the characteristic ``Y'' shape for oblique modes
or a vertical line at the spanwise limit \cite{schmid_stability_2001}, with the damped Squire and Orr-Sommerfeld
modes being nearly degenerate. Odd viscosity hybridizes the two families of modes and lifts the degeneracy.
The unstable mode that arises for the lowest value of $\rm{Ro}$ in (a) corresponds to an eigenvalue with
real part (phase velocity) $c_r\approx0.92$, and the one in (b) to an eigenvalue with $c_r\approx0.58$.
\label{fig:eigval plot}}
\end{figure}
\par\end{center}

\subsection{Dependence of the instability on even-to-odd viscosity ratio}
\label{Dependence}
\begin{figure}[H]
\begin{centering}
\includegraphics[scale=0.53]{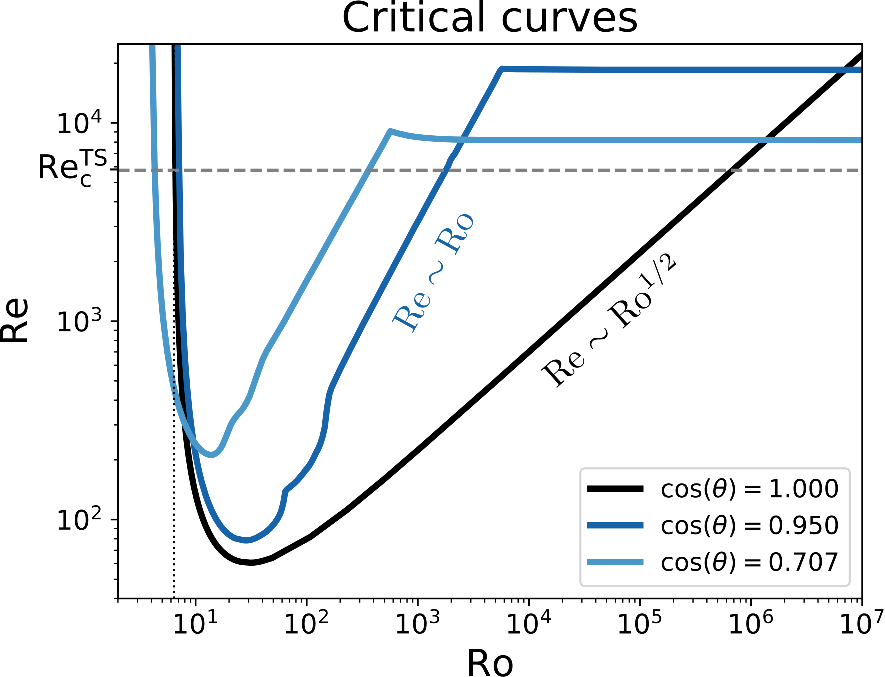}
\par\end{centering}
\caption{Critical Reynolds number as a function of odd Reynolds number for
various wave-vector angles $\theta$. The curves are obtained by scanning
all values of $k$ for a given point $\left({\rm Re},{\rm Ro},\theta\right)$
and extracting the minimal ${\rm Re}$ for which instability occurs.
The gray dashed horizontal line indicates ${\rm Re}_{\rm c}=5772$, the critical
Reynolds number for Poiseuille flow in a non-odd fluid with a streamwise
perturbation wavevector, $\left(\alpha,\beta\right)=\left(k,0\right)$.
The vertical dotted line indicates the analytic limit
$\rm{Ro}_{\rm c}\approx6.412$ where odd viscosity 
completely restabilizes all spanwise modes.
\label{fig:Critical curves Re(Ro)}}
\end{figure}

We now investigate the dependence of the critical Reynolds number ${\rm Re}$ on
the value of odd viscosity 
(Fig. \ref{fig:Critical curves Re(Ro)}).
For each fixed wavevector angle $\theta=\arctan\left(\beta/\alpha\right)$, 
we determine the minimum Reynolds number at which an instability first appears.


As discussed above, the odd-viscosity-induced instability emerges first
for spanwise perturbations ($\theta=\pi/2$), whereas streamwise
perturbations ($\theta=0$) are unaffected by odd viscosity. We first
consider the generic case $0 < \theta < \pi/2$. 
In the absence of odd viscosity,
the critical Reynolds number is determined by the onset of the
classic Tollmien--Schlichting instability. By Squire's
transformation (Appendix \ref{appendix subsec Squire transformation}),
it is given by
\begin{equation}
{\rm Re_{c}^{{\rm TS}}}(\theta)=\frac{{\rm Re}_{{\rm c}}^{{\rm TS}}(\theta=0)}{\cos\theta}\approx\frac{5772.2}{\cos\theta}.\label{eq:Critical Re Squire transformation}
\end{equation}
This classic value corresponds to a ${\rm Ro\rightarrow\infty}$
plateau of the non-spanwise ($\theta\neq\pi/2$) curves in Fig. \ref{fig:Critical curves Re(Ro)}.
To  the left of the cusp  
the instability with the lowest critical Reynolds number is the odd-viscosity-induced instability.
Near the cusp, the critical Reynolds number decreases linearly with ${\rm Ro}$, i.e.,  ${\rm Re}\sim{\rm Ro}$.
The linear dependence
persists down to remarkably small  critical Reynolds numbers
(${\rm Re}\approx100$ for $\sin\theta=0.95)$. For lower values of ${\rm Ro}$, odd
viscosity has a stabilizing effect and the critical value
of ${\rm Re}$ increases. Eventually the flow becomes linearly stable for all Reynolds numbers.

The behavior of spanwise perturbations ($\theta=\pi/2$) is distinctly
different from the generic case of oblique perturbations ($\theta\neq\pi/2$). In the absence of
odd viscosity, such perturbations are always damped. However, for any
small but non-zero odd viscosity, unstable spanwise
modes arise above a critical
Reynolds number, as shown by the black curve in
Fig.~\ref{fig:Critical curves Re(Ro)}. 
Unlike the generic case, where ${\rm Re} \sim {\rm Ro}$, 
the spanwise case obeys the scaling ${\rm Re} \sim {\rm Ro}^{1/2}$
for weak to moderate odd viscosity (${\rm Ro}\gtrsim100$).
For stronger odd viscosity, the trend reverses and odd viscosity becomes stabilizing.
The critical Reynolds number increases as ${\rm Ro}$ decreases, and the instability
eventually disappears at ${\rm Ro}\approx6.412$. A more detailed
analysis of the spanwise case will be given in the next section.

The qualitative difference in behavior between oblique and spanwise modes is due to the latter
not being convected by the base flow (being uniform in the streamwise direction).
The oblique modes develop critical layers at the planes in which the base flow
velocity matches the streamwise phase velocity of the perturbation.
In the vicinity of the critical
layer, the components of the perturbation develop large gradients
(scaling as $\left(\alpha{\rm Re}\right)^{\gamma}$,
with $\gamma$ being a positive rational number that depends on 
the location of the critical layer \citep{drazin_hydrodynamic_2004}). 
Outside the critical layers and away from the walls, viscous forces (even and odd)
are negligible compared to momentum transfer by convection.
Therefore, for spanwise modes, viscous forces are significant throughout the whole channel,
whereas for oblique modes they are concentrated in the vicinity of thin layers,
resulting in different scaling behavior for the critical curves.

We note that the critical curve for the spanwise-mode instability closely resembles the
one found in rotating shear flows, where the Coriolis force
acts on the fluid in the rotating frame of reference 
\citep{lezius_roll-cell_1976,alfredsson_instabilities_1989,wall_nonlinear_2006,brethouwer_stability_2025}.
The two problems share several key features, including the scaling  ${\rm Re} \sim {\rm Ro}^{1/2}$, 
the non-monotonic shape of the critical curve,  and the eventual stabilization in the limit of large odd viscosity (or rapid rotation).
The resemblance traces back to the analogy between the odd viscous force 
[the last term in Eq.~(\ref{eq:N-S full}) of Appendix \ref{appendix oss-derivation}]
and a wavenumber-dependent Coriolis force, with the wavenumber dependence arising from the Laplacian.

Despite the above-mentioned similarities between the two problems, there is a key difference.
In the rotating-fluid case, the instability is exclusive
to spanwise modes, and a slight deviation from $\theta=\pi/2$ renders
the modes stable.  By contrast, the  odd-viscosity-induced instability
is more robust and persists for $\theta\neq \pi/2$.
This robustness is due to the dependence of the odd viscous force on spatial derivatives of the velocity field.
As mentioned above, oblique modes develop critical layers with large velocity gradients.
Because the odd viscous force scales with the Laplacian of the velocity field,
it can compete with the dissipative viscous force (which is also large at the critical layer).
This enables the odd-viscosity-induced instability to occur also for
modes with $\alpha\neq 0$, in contrast to the case of a rotating fluid,
where a small Coriolis force remains negligible throughout the entire channel.

\subsection{Spanwise modes -- origin of the odd viscous instability}
\label{spanwise}
Since the instability threshold is reached first by spanwise-modulated modes, it is instructive to analyze this case analytically.
For $\alpha=0$ and $\beta=k$, the odd Orr--Sommerfeld--Squire
system [Eq. (\ref{eq:OS-SQ odd matrix operator})] simplifies. 
One can obtain a single equation for the wall-normal velocity,
\begin{equation}
\left[\left(\omega-\frac{i}{{\rm Re}}\Delta_{k}\right)^{2}\Delta_{k}-\frac{\beta^{2}}{{\rm Ro}}\Delta_{k}\left(V^{\prime}-\frac{1}{{\rm Ro}}\Delta_{k}\right)\right]u_{y}=0,\label{eq:spanwise limit}
\end{equation}
supplemented with the boundary conditions
\begin{align}
u_{y}\left(\pm1\right) & =\left[\frac{\partial}{\partial y}u_{y}\right]_{y=\pm1}\nonumber \\
= & \left[\left(\Delta_{k}+i\omega{\rm Re}-\left(\frac{\beta{\rm Re}}{{\rm Ro}}\right)^{2}\right)\Delta_{k}u_{y}\right]_{y=\pm1}=0\label{eq:spanwise limit BC}
\end{align}
Here, the effect of the base flow on the dynamics is manifested only
in the vortex-tilting term containing the base-flow vorticity $\Omega_{z}=-V^{\prime}$;
there are no convection terms, since the perturbation is uniform along
the streamwise direction. The eigenmodes thus do not possess the critical
layer that develops for modes with $\alpha\neq 0$. 

The different roles played by odd viscosity can be clearly
understood from the last two terms in Eq. (\ref{eq:spanwise limit}).
The $\sim1/{\rm Ro^{2}}$ term corresponds
to the restoring, spring-like effect of odd viscosity \citep{kirkinis_taylor_2023}.
The term proportional to $V^{\prime}/{\rm Ro}$ is due to vortex-tilting
of the base-flow vorticity into the wall-normal direction (responsible
for the lift-up effect, as elaborated on in the next section),
and the odd viscous coupling of the wall-normal vorticity
back to the wall-normal velocity. It is this term
that allows energy extraction
from the base flow, thereby enabling instability.
Lastly, the $\sim1/{\rm Re}^{2}$ term represents the viscous dissipation acting on the wall-normal velocity and vorticity.
At small values of odd viscosity, balance between the viscous dissipation and the $\sim1/\rm{Ro}$ energy-extracting term
leads to the ${\rm Re}\sim{\rm Ro}^{1/2}$
scaling of the critical Reynolds number for spanwise waves (right-side part of the black curve in Fig. \ref{fig:Critical curves Re(Ro)}).
At large values of odd viscosity, the restoring $\sim 1/{\rm Ro^{2}}$ term dominates the energy-extracting term,
resulting in the restabilization of the system.


By numerically solving the eigenvalue problem of Eq. (\ref{eq:spanwise limit}),
we find that the transition between stability and instability always occurs by
the least stable eigenvalue crossing through the origin.
Thus, the marginal perturbation at the stability threshold
corresponds to a stationary mode with $\omega=0$.
This behavior is consistent with the principle of exchange of stabilities (PES).
While our analysis is numerical, the PES has been proven analytically for
the related problems of Rayleigh--B\'enard convection and rotating shear flows
\citep{chandrasekhar_hydrodynamic_1961,lezius_roll-cell_1976,herron_principle_2001,herron_principle_2003}.  
We therefore expect that a similar analytical proof can be constructed for the present problem.

Assuming the validity of the PES,
we are able to derive several analytic results for the
marginal stability curve due to Eq. (\ref{eq:spanwise limit})
(Appendix \ref{appendix analytics}).
In the inviscid (dissipationless) ${\rm Re}\rightarrow \infty$ limit,
after substituting $\omega=0$, it can be seen that Eq. (\ref{eq:spanwise limit}) approaches the limit
of the Airy equation (with $V^\prime=-2y$ for Poiseuille flow), with some care required
to account for the boundary-layer structure of the equation.
This leads to the minimal odd Reynolds number (maximal odd viscosity)
$\rm{Ro}_{\rm{c}} \approx 6.412$ for the spanwise instability
(vertical dotted line in Fig. \ref{fig:Critical curves Re(Ro)}),
corresponding to the restabilization threshold.

Lastly, we mention that further analytic progress for oblique
(non-spanwise) modes (for instance, deriving the scaling of the critical curves for that case)
is more complicated and remains to be done.
We expect it to require a WKB analysis of the Orr-Sommerfeld-Squire system of equations,
as was done in Ref. \cite{chapman_subcritical_2002} for the case without odd viscosity.
Nonetheless, we argue that the competition between the energy-extracting and restoring roles
of odd viscosity generalizes to all modes, resulting in the non-monotonic behavior
of the critical curves in Fig. \ref{fig:Critical curves Re(Ro)}.
The principle of exchange of stabilities is particular to the spanwise limit.

\subsection{Mechanism of the odd-viscosity-induced instability: from transient
to exponential growth}
\label{mechanism}

Having established that odd viscosity qualitatively changes the stability diagram, we now capitalize on this problem to address two broader questions. 
First, can one determine, without explicitly solving the stability problem, whether an energy-conserving perturbation can induce an instability? 
Second, what is the specific physical mechanism by which odd viscosity destabilizes shear flow?

\subsubsection{Instability induced by energy-conserving perturbations: mathematical perspective}
\label{instability mathematics}

The ability of an energy-conserving term (here, odd viscosity) to induce an instability of the
Orr--Sommerfeld--Squire operator is rooted in the non-normal nature of the unperturbed operator \cite{jose_non_normal_2020}
\footnote{We note Ref. \cite{jose_non_normal_2020} for a related discussion of the connection between rotating-flow 
instabilities and the pseudospectrum of the Navier--Stokes operator.}.
An operator is non-normal if it does not commute with its adjoint, and its eigenmodes are generally non-orthogonal. 
Consequently, modal stability does not fully characterize the dynamics: even when all eigenmodes are stable, 
suitable superpositions of them may exhibit transient growth before eventually decaying.
In shear flows, this transient growth can lead to a substantial amplification of the perturbation energy.
Here we will show that it is the presence of transiently growing disturbances (in the original operator)
that enables energy-conserving perturbations to induce an instability. The energy-conserving perturbation
can do so by promoting a transiently growing disturbance into a self-sustaining, true unstable eigenmode.

To demonstrate this general phenomenon, let us consider the time evolution of a state $\vert \phi(t) \rangle$
[here we use the bra-ket notation of quantum mechanics, the states being understood as functions in some space, e.g.,
vectors with their two components being functions of $y \in \left[ -1, 1 \right]$ in the Orr--Sommerfeld--Squire case, Eq. (\ref{eq:OS-SQ odd matrix operator})], governed by
\begin{equation}
\frac{\partial}{\partial t} \vert \phi(t) \rangle =-iS \vert \phi(t) \rangle,\label{eq:time evolution}
\end{equation}
where the time-evolution operator $S=S_{0}+S_{H}$ consists of an
unperturbed part 
$S_{0}$ and a Hermitian perturbation $S_{H}=S_{H}^{\dagger}$.
Note that Hermiticity (and the normality of an operator) is defined with respect
to a chosen inner product, demanding $\langle  S_{H} \phi \vert \psi \rangle = \langle \phi \vert S_{H} \psi \rangle$
for all pairs of states. In the context of our problem, it is useful to take
the inner product in which the squared norm $\left\Vert \phi\right\Vert ^2 =\left\langle \phi \vert \phi \right \rangle$
corresponds to the energy; we will hereafter refer to it as the energy-norm inner product
(see Appendix \ref{appendix Non-normal operators} for the definition of the energy-norm
inner product and further derivations).

The instantaneous growth rate of a state (i.e., the growth of its squared norm $\left\Vert \phi\right\Vert ^2 =\left\langle \phi \vert \phi \right \rangle$)
is determined only by the anti-Hermitian part of $S$, 
\begin{equation}
\frac{d\left\Vert \phi\right\Vert ^{2}}{dt}= -i \left\langle \phi\vert\left(S-S^{\dagger}\right)\phi\right\rangle = -i \left\langle \phi\vert\left(S_{0}-S_{0}^{\dagger}\right)\phi\right\rangle .
\end{equation}

The conservation of the norm by Hermitian operators
is the motivation for choosing the energy-norm inner product in our case;
this choice makes Hermiticity equivalent to energy conservation.
Now, consider the case where the Hermitian perturbation induces an instability,
i.e., the full operator $S$ has an unstable eigenmode $\vert \phi_{{\rm unstable}} \rangle$
while the base operator $S_{0}$ has only stable eigenmodes. Instability
implies that an initial perturbation $\vert \phi(t=0) \rangle=\vert \phi_{{\rm unstable}} \rangle$ grows in norm
when it evolves under Eq. (\ref{eq:time evolution}).
Since the growth
rate is controlled only by the anti-Hermitian part of $S$, it must
have been positive to begin with, i.e., the same state would undergo (momentary) growth also
when evolving under the original operator $S_{0}$. Therefore, if the perturbed operator $S$ has an unstable eigenmode,
the unperturbed operator $S_0$ must have transiently growing disturbances.

The above discussion can be succinctly reformulated in terms of the modern theory of pseudospectra,
using the concept of the numerical range \citep{reddy_pseudospectra_1993,schmid_nonmodal_2007}.
The numerical range of an operator $S$ is defined as the set of all possible expectation
values for unit-norm functions (with the inner product of interest here still being the energy-norm inner product),


\begin{equation}
\label{range}
F(S)=\left\{
\langle\phi|S\phi\rangle:\|\phi\|=1
\right\}.
\end{equation}

The maximal extent of the range along the imaginary axis is known as the
numerical abscissa,
\begin{equation}
\label{eq:abscissa}
w(S)=\max \operatorname{Im} F(S).
\end{equation}
When $w(S)>0$, the numerical range extends above the real axis, and
disturbances with a positive instantaneous growth rate exist
\citep{schmid_nonmodal_2007}. As can be seen from its definition,
the numerical range contains all the eigenvalues of the operator.
Nonetheless, for non-normal operators, the numerical abscissa can extend
into the unstable region $w>0$ even when all eigenvalues are below the real axis.

As follows immediately from the definitions [Eqs. (\ref{range}) and (\ref{eq:abscissa})], a
Hermitian perturbation leaves the numerical abscissa invariant,
\begin{equation}
w(S)=w(S_0).
\end{equation}
Nevertheless, the perturbation can shift the eigenvalues, whose imaginary parts remain
bounded by the numerical abscissa,
\begin{equation}
\operatorname{Im}\lambda(S)\leq w(S)=w(S_0).
\end{equation}

Thus, a Hermitian perturbation can induce an instability in a stable
system only if the numerical range of the unperturbed operator extends
above the real axis, $w(S_0)>0$, implying that the unperturbed system
supports transient growth. If $w(S_0)\leq0$, neither transient growth
nor an instability induced by a Hermitian perturbation is possible.

\begin{figure}
\begin{centering}
\includegraphics[scale=0.48]{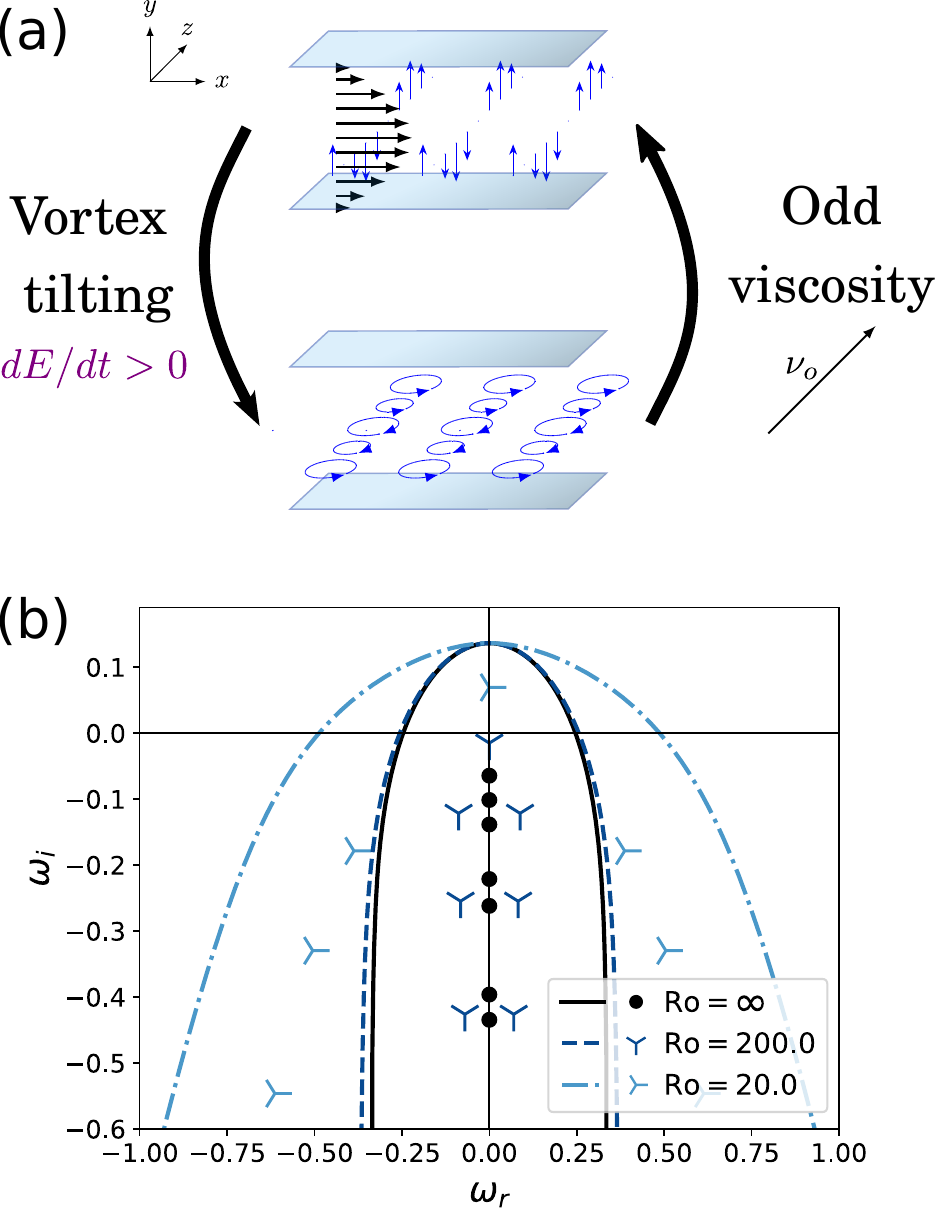} 
\par\end{centering}
\caption{(a) Mechanism of the odd-viscosity-induced instability. In shear flows,
perturbations with spatial variation parallel to the base flow vorticity
extract energy from the base flow through the vortex-tilting mechanism,
where the wall-normal velocity acts as a forcing term for the wall-normal
vorticity. In conventional fluids, the forcing velocity eventually
decays, and subsequently, the whole perturbation; hence, the perturbation
is transiently growing. Odd viscosity couples the normal vorticity
with the normal velocity, turning the perturbation into an exponentially
growing mode. (b) Numerical
range and eigenvalues for ${\rm Re}=100$, $\alpha=0$, $\beta=2$,
and varying values of ${\rm Ro}$ ($\omega=\omega_r + i \omega_i$).
Eigenvalues are bounded
within the range. Odd viscosity does not change
the extent of the range along the imaginary axis, a consequence of
its energy-conserving nature. \label{fig:mechanism and range}}
\end{figure}

Coming back to our problem, we conclude that the odd-viscosity-induced instability may exist
only at parameters ($\alpha$, $\beta$, ${\rm Re}$) for which the numerical
range of the conventional (without odd viscosity) Orr--Sommerfeld--Squire
operator extends above the real axis.
In Fig. ~\ref{fig:mechanism and range}(b), we plot the numerical range
and eigenvalues of the Orr--Sommerfeld--Squire operator at different
values of odd viscosity, for fixed Reynolds number and wave numbers.
As ${\rm Ro}$ is varied, the imaginary extent of the numerical range
remains unchanged, while the eigenvalues shift and eventually enter the
unstable half-plane at sufficiently strong odd viscosity.

We emphasize that the above discussion applies generally to
energy-conserving perturbations of non-normal operators. In the present
work, odd viscosity plays the role of an energy-conserving addition to
the linearized Navier--Stokes operator. Another closely related example
is provided by rotating flows, where the Coriolis force acts as an
energy-conserving term that can induce instability
\citep{lezius_roll-cell_1976}.

\subsubsection{Transient mode conversion: effects of odd viscosity on lift-up disturbances}

The above analysis establishes transient growth in the unperturbed
system as a necessary, but not sufficient, condition for an
energy-conserving perturbation to induce an instability.
We now turn to the specific physical mechanism of the
odd-viscosity-induced instability. We show how odd viscosity modifies the dynamics of
disturbances that undergo transient growth via the lift-up
effect, providing a feedback mechanism that can turn the transient
growth into an exponentially growing instability.

As discussed above, the odd-viscosity-induced instability first emerges
for spanwise disturbances. In conventional shear flows, such disturbances
are particularly susceptible to transient growth through the lift-up
mechanism \citep{butler_three-dimensional_1992}. A wall-normal velocity
perturbation $u_y$ displaces fluid across the base shear, producing a
large streamwise velocity perturbation, or streak. In the present
formulation, this corresponds to the generation of wall-normal vorticity
$\varpi_y$, whose amplitude can grow by a factor proportional to
${\rm Re}$. In a conventional fluid this coupling is one-way: $u_y$
generates $\varpi_y$, but there is no corresponding feedback from
$\varpi_y$ to $u_y$. Consequently, the growth is transient and is
eventually followed by exponential decay.

Odd viscosity provides precisely this missing feedback. The induced
wall-normal vorticity generates an odd-viscous force in the wall-normal
direction, coupling $\varpi_y$ back to $u_y$
[Fig.~\ref{fig:mechanism and range}(a)]. This closes the feedback loop
and allows the transient lift-up growth to become a self-sustaining,
exponentially growing instability.

In view of the previous section, the region of the numerical range above
the real axis corresponds to disturbances susceptible to instantaneous
growth. Remarkably, odd viscosity couples precisely to these disturbances,
turning their transient growth into a true exponential instability.

\section{Conclusions}

Our results demonstrate that odd viscosity can act as a genuine
instability mechanism in shear flows. For plane Poiseuille flow, it
renders a large part of the parameter space unstable and reduces the
critical Reynolds number from the conventional value
${\rm Re}_c \approx 5772$ to values as low as ${\rm Re}_c \approx 62$.
The instability originates from the non-normal dynamics of the
conventional Orr--Sommerfeld--Squire system, which enable transient growth.
One of the pathways to transient growth is through the lift-up mechanism, in which
a disturbance in the wall-normal velocity generates wall-normal vorticity, 
extracting energy from the base flow in the process.
Odd viscosity provides the reverse coupling from the
wall-normal vorticity back to the wall-normal velocity.
It thereby closes the feedback loop and converts transient
growth into a self-sustaining, exponentially growing instability.

More generally, our results establish a connection between systems governed by
a non-normal operator and instabilities induced by energy-conserving perturbations.
For non-normal operators, it is useful to consider the numerical range,
which determines the potential for transiently growing disturbances and
constrains the operator's eigenvalues. Energy-conserving perturbations to an operator
do not change the projection of the numerical range along the imaginary axis.
This implies that such perturbations cannot generate an instability unless
the numerical range of the
unperturbed operator already extends into the unstable half-plane.
Thus, transient growth in the pristine system is a necessary
condition for destabilization by an energy-conserving perturbation
to the time-evolution operator. 

Stable systems that permit transient growth thus allow the seemingly paradoxical scenario,
in which adding an energy-conserving coupling can lead to exponential energy growth.
As exemplified here, this is possible when the coupling enables continuous
energy extraction through the preexisting growth pathway.
Odd viscosity provides a concrete hydrodynamic realization of this
general mechanism, with the Coriolis force in rotating shear flows
providing another closely related example.

The unusually low critical Reynolds numbers found here may have
important implications for the transition to turbulence. In
conventional plane Poiseuille flow, transition is observed well below
the linear instability threshold \cite{carlson_flow_visualization_1982}, in a regime where transient growth
and nonlinear effects play an essential role. Odd viscosity converts
precisely this type of transient amplification into linear
instability, potentially extending the regime in which modal
instability controls the onset of complex flow dynamics.
Odd viscosity has also recently been shown to profoundly modify
nonlinear energy transfer in fully developed turbulence, arresting
turbulent cascades and producing scale selection through odd waves
\citep{de_wit_pattern_2024}. An interesting open question is how the
linear instability identified here evolves into the nonlinear regime
and whether its unstable modes influence the ensuing turbulent
cascade.

Furthermore, the lowering of the critical Reynolds number even below the
threshold for subcritical transition (where turbulence can be self-sustained
due to transient growth and nonlinearities)
may allow to observe turbulent behavior in hydrodynamic systems where
the high Reynolds number regime is currently inaccessible, such as
viscous electronic flows \cite{lucas_hydrodynamics_2018,fritz_hydrodynamic_2024,hui_hydrodynamics_2025}.
In these systems, reaching a high Reynolds number is challenging
due to heating effects at strong driving and sample impurities.
However, recent experiments in rhombohedral graphene \cite{zhang_imaging_2026}
have reached an estimated Reynolds number of $\mathrm{Re}\approx 50$. 
Furthermore, time-reversal symmetry breaking has been observed in
these materials \cite{zhou_half_2021},
permitting finite odd viscosity. Thus, the odd-viscosity-induced instability
may be experimentally relevant to these systems,
bringing turbulent behavior in viscous electron hydrodynamics
within experimental reach.

A further aspect of our results concerns non-Hermitian dynamics in open systems.
We find that the odd Orr–Sommerfeld–Squire operator exhibits numerous exceptional
points (EPs) over a broad parameter range.
EPs are points of degeneracy where the eigenmodes coalesce simultaneously
with the eigenvalues, and they are a unique feature of non-Hermitian systems
\cite{berry_physics_2004,heiss_exceptional_2004,xue_essay_2026}.
Beyond mathematical significance,
EPs in odd shear flows may be experimentally detectable,
via probing the system along a closed path in the parameter space,
changing the parameters adiabatically.
Odd viscosity provides a simple knob for periodic modulation of the system,
since it can be controlled externally; for example, when induced by an external
magnetic field, it can be modulated by varying the field.
Mathematically, to make a closed non-trivial path, the second varying parameter
can be any of the wavenumber components or the Reynolds number.
Experimentally, it seems more feasible to vary the Reynolds number,
for example by controlling the viscosity via its temperature dependence.
The base flow profile needs to be kept constant, which may be done by
forcing flow at fixed velocity in the center of the channel.

Finally, the connection between transient growth, numerical range, and
instability suggests a strategy for stability analysis  that extends beyond hydrodynamics.
For a non-normal dynamical system, one can first identify  the
disturbances associated with the upper-half-plane part of the numerical range and
then determine which energy-conserving perturbations couple strongly
to them. This provides a way of identifying potentially destabilizing
perturbations without solving the full stability problem separately
for every perturbation.

\begin{acknowledgments}
We thank Y. Avron, G. Falkovich, D. Kessler, and A. Peer for stimulating and insightful discussions.
\end{acknowledgments}

\begin{widetext}

\appendix
\renewcommand{\theequation}{\thesection\arabic{equation}}

\section{Derivation of the odd Orr--Sommerfeld--Squire equations \label{appendix oss-derivation}}

In this Appendix, we show the derivation of the odd Orr--Sommerfeld--Squire equation, Eq. 
(\ref{eq:OS-SQ odd matrix operator}) of the main text.
We follow the standard linearization procedure, but take into account the odd viscous terms. 
We start with the Navier--Stokes (NS) equations for an incompressible fluid,

\begin{align}
\frac{\partial v_{i}}{\partial t}+\left(v_{j}\partial_{j}\right)v_{i} & =-\frac{1}{\rho}\partial_{i}P+\frac{1}{\rho} \partial_{j}\sigma_{ij},\label{eq:N-S general}\\
\partial_{i}v_{i} & =0, \label{eq:incompressibility}
\end{align}
where $\sigma_{ij}$ is the viscous stress tensor,

\begin{equation}
\sigma_{ij}=\eta_{ijkl} \frac{\partial v_{k}}{\partial x_{l}}.
\end{equation}
The viscosity tensor $\eta$ can be decomposed into its even and odd parts,

\begin{align}
\eta_{ijkl} & =\eta_{ijkl}^{{\rm e}}+\eta_{ijkl}^{{\rm o}},\\
\eta_{ijkl}^{{\rm e}} & =\eta_{klij}^{{\rm e}},\\
\eta_{ijkl}^{{\rm o}} & =-\eta_{klij}^{{\rm o}}.
\end{align}
For the even part, we take the conventional shear viscosity,

\begin{equation}
\eta_{ijkl}^{{\rm e}}=\eta^{{\rm e}}\left(\delta_{ik}\delta_{jl}+\delta_{il}\delta_{jk}\right).
\end{equation}

In the presence of cylindrical symmetry, there may be 8 independent
coefficients for the odd part of the viscous tensor \citep{khain_stokes_2022}.
For simplicity, in this work we focus on one combination
($\eta_o=2\eta_4$ in the notations of Ref. \cite{kirkinis_evanescent_2024}, also used
in Refs. \cite{markovich_odd_2021,khain_stokes_2022}).
Choosing the distinguished axis as the $z-$axis, the
odd part of the viscosity tensor is given by

\begin{equation}
\eta_{ijkl}^{{\rm o}}=\frac{\eta^{{\rm o}}}{2}\left(\delta_{ik}\epsilon_{jlz}+\delta_{jl}\epsilon_{ikz}+\delta_{jk}\epsilon_{ilz}+\delta_{il}\epsilon_{jkz}\right).\label{eq:odd viscosity component}
\end{equation}
The NS equations (\ref{eq:N-S general}) then read [defining the
kinematic viscosities as $\nu\equiv\eta/\rho$
and utilizing the incompressibility condition (\ref{eq:incompressibility})]

\begin{equation}
\frac{\partial v_{i}}{\partial t}+\left(v_{j}\partial_{j}\right)v_{i}=-\frac{1}{\rho}\partial_{i}\tilde{P}+\nu^{{\rm e}}\triangle v_{i}+\frac{\nu^{{\rm o}}}{2}\epsilon_{ijz}\triangle v_{j},\label{eq:N-S full}
\end{equation}
where a part of the odd viscous stress modifies the pressure according
to

\begin{equation}
\tilde{P}=P - \frac{\rho \nu^{{\rm o}}}{2} \epsilon_{jkz}\partial_{j}v_{k}.
\end{equation}
The pressure can be eliminated by taking the curl of Eq. (\ref{eq:N-S full}), yielding
the NS equations for the vorticity $\boldsymbol{\Omega}=\boldsymbol{\nabla}\times\boldsymbol{v}$,

\begin{equation}
\frac{\partial\Omega_{i}}{\partial t}+\left(v_{j}\partial_{j}\right)\Omega_{i}=\left(\Omega_{j}\partial_{j}\right)v_{i}+\nu^{{\rm e}}\triangle\Omega_{i}+\frac{\nu^{{\rm o}}}{2}\partial_{z} \triangle v_{i}.
\end{equation}

To study the stability of a given base flow, we write the velocity
and (modified) pressure fields as steady-state solutions plus perturbative
parts,

\begin{align}
\boldsymbol{v} & =\boldsymbol{V}+\boldsymbol{u},\\
\tilde{P} & =\tilde{P}_{0}+\tilde{p}.
\end{align}
Linearizing the NS equations (\ref{eq:N-S full}) with respect to the perturbation, we obtain

\begin{align}
\frac{\partial u_{i}}{\partial t}+\left(V_{j}\partial_{j}\right)u_{i}+\left(u_{j}\partial_{j}\right)V_{i} & =-\partial_{i}\tilde{p}+\frac{1}{{\rm Re}}\triangle u_{i}+\frac{1}{{\rm Ro}}\epsilon_{ijz}\triangle u_{j},\label{eq:linearized N-S}\\
\partial_{i}u_{i} & =0,\label{eq:incompressible flow}
\end{align}
where velocities and lengths are made dimensionless by dividing by the
typical scales $\bar{V}$ and $\delta$, and the even and odd Reynolds numbers
are defined by ${\rm Re=\bar{V}\delta/\nu^{{\rm e}}}$ and ${\rm Ro=2 \bar{V}\delta/\nu^{{\rm o}}}$.

Let us now consider a steady-state shear flow, where the base velocity is of the
form $\boldsymbol{V}=V(y)\hat{x}$. We perform a modal expansion in
time and the spatial streamwise ($x$) and spanwise ($z$) axes,

\begin{align}
\boldsymbol{u} & =\boldsymbol{u}(y)\exp\left[i\left(\alpha\left(x-ct\right)+\beta z\right)\right],\label{eq:u perturbation}\\
\tilde{p} & =\tilde{p}(y)\exp\left[i\left(\alpha\left(x-ct\right)+\beta z\right)\right].\label{eq:p perturbation}
\end{align}
Note that in this expansion, we have chosen to denote by $c$ the
phase velocity along the streamwise direction. 

Linearizing the flow around the base shear flow, $\boldsymbol{v}=\left(u_{x}+V(y),u_{y},u_{z}\right)$
and $\boldsymbol{\Omega}=\left(\varpi_{x},\varpi_{y},\varpi_{z}-V^{\prime}\right)$,
the corresponding linearized equations for $\varpi_{z}$ and $\varpi_{y}$
read

\begin{align}
L_{{\rm SQ}}^{c}\left(i\alpha u_{y}-\frac{\partial u_{x}}{\partial y}\right)+i\beta V^{\prime}u_{z}-V^{\prime\prime}u_{y}-i\beta\frac{1}{{\rm Ro}}\Delta_{k}u_{z} & =0,\label{eq:first eq. in linearized vorticity eqs}\\
L_{{\rm SQ}}^{c}\varpi_{y}+i\beta\left[V^{\prime}-\frac{{\rm 1}}{{\rm Ro}}\Delta_{k}\right]u_{y} & =0,\label{eq:omega_y linearized eq}
\end{align}
with

\begin{equation}
L_{{\rm SQ}}^{c}=i\alpha\left(V-c\right)-\frac{1}{{\rm Re}}\Delta_{k}
\end{equation}
(note the different notation compared to $L_{{\rm SQ}}$ defined in the main text; the relation is $L_{{\rm SQ}}^{c}=L_{{\rm SQ}}-i\alpha c$).
Eq. (\ref{eq:omega_y linearized eq}) appears as the bottom row of
the matrix equation (\ref{eq:OS-SQ odd matrix operator}) in the main text.
To get the top row of  Eq. (\ref{eq:OS-SQ odd matrix operator}), 
we utilize the incompressibility condition
and the definition of the wall-normal vorticity,

\begin{align}
\frac{\partial u_{y}}{\partial y} & =-i\left(\alpha u_{x}+\beta u_{z}\right),\label{eq:du_y_dy with u_x u_z}\\
\varpi_{y} & =i\left(\beta u_{x}-\alpha u_{z}\right),\label{eq:eta with u_x u_z}
\end{align}
and express $u_{x}$ and $u_{z}$ by 

\begin{align}
u_{x} & =\frac{i}{k^{2}}\left(\alpha\frac{\partial u_{y}}{\partial y}-\beta\varpi_{y}\right),\label{eq:u_x with u_y omega_y}\\
u_{z} & =\frac{i}{k^{2}}\left(\beta\frac{\partial u_{y}}{\partial y}+\alpha\varpi_{y}\right).\label{eq:u_z with u_y omega_y}
\end{align}
Substituting Eqs. (\ref{eq:u_x with u_y omega_y}) and (\ref{eq:u_z with u_y omega_y})
in Eq. (\ref{eq:first eq. in linearized vorticity eqs}), we obtain



\begin{equation}
i\alpha L_{{\rm SQ}}^{c}\Delta_{k}u_{y}+\beta^{2}V^{\prime}\frac{\partial u_{y}}{\partial y}+k^{2}V^{\prime\prime}u_{y}-i\beta\frac{\partial}{\partial y}\left(L_{{\rm SQ}}^{c}\varpi_{y}\right)-\frac{\beta}{{\rm Ro}}\Delta_{k}\left(\beta\frac{\partial u_{y}}{\partial y}+\alpha\varpi_{y}\right)=0.
\end{equation}
Substituting $L_{{\rm SQ}}^{c}\varpi_{y}$ from Eq. (\ref{eq:omega_y linearized eq})
and simplifying, we arrive at the first row of Eq. (\ref{eq:OS-SQ odd matrix operator})  in the main text.

\section{Solutions of the Orr--Sommerfeld--Squire equations in
the even fluid limit \label{appendix classic OSS eqs}}
To make the paper self-contained, we show how in the conventional case
($1/\rm{Ro}=0$) the OSS equation (\ref{eq:OS-SQ odd matrix operator})
acquires a triangular structure,
identify the families of its solutions
and discuss their main properties \citep{drazin_hydrodynamic_2004,schmid_stability_2001}.
We also discuss the parity symmetry of the equations
in the case of a symmetric base flow, which is reduced by odd viscosity to an extended symmetry.

\subsection{Orr-Sommerfeld and Squire modes}

The conventional Orr--Sommerfeld--Squire equations [corresponding
to taking $1/{\rm Ro}=0$]
in Eq. (\ref{eq:OS-SQ odd matrix operator}) possess a triangular
structure, allowing one to separate the solutions into two separate
families, Orr-Sommerfeld  (OS) and Squire modes.
The OS modes are obtained by first solving the Orr-Sommerfeld equation
$\left( L_{{\rm OS}} +i\alpha c \Delta_k \right) u_{y}=0$. The solutions for the wall-normal velocity
$u_{y}$ then serve as a source term in the equation for the wall-normal
vorticity $\varpi_{y}$ (the source term corresponding to the vortex-tilting
mechanism). The OS modes are the modes responsible for the linear
instability in conventional fluids (the Tollmien-Schlichting instability).

The second family of modes, the Squire modes, are solutions with vanishing
wall-normal velocity $u_{y}=0$ and a wall-normal vorticity that solves
the homogeneous equation $\left( L_{{\rm SQ}} - i\alpha c \right) \varpi_{y}=0$. In conventional fluids,
the Squire modes are always damped, as can be shown by a simple
energetic analysis \citep{schmid_stability_2001}. However, they are relevant for
transient growth.

\subsection{Squire transformation \label{appendix subsec Squire transformation}}

In a conventional  fluid, any 3D OS mode with wave vector $\boldsymbol{k}=\alpha\hat{k}_{x}+\beta\hat{k}_{z}$
can be mapped to an equivalent 2D OS mode, using the transformation

\begin{equation}
\alpha\rightarrow k,\qquad {\rm Re} \rightarrow\alpha  {\rm Re} /k.\label{eq:Squire transformation}
\end{equation}
Therefore, each 3D OS mode corresponds to a 2D one with a lower Reynolds
number. It thus follows that the 3D instabilities always arise later
(at a higher critical Reynolds number) than the 2D ones. This statement
is known as Squire's theorem. Together with the fact that the exclusively
3D Squire modes are always damped, this leads to the conclusion that
for modal instabilities, it is sufficient to consider only 2D perturbations
in conventional fluids \citep{squire_stability_1933}.

It is worth mentioning that the Squire transformation has implications for
the scaling and optimization of the transient growth in the initial-value problem;
see Ref. \cite{john_soundar_jerome_extended_2014} for a discussion and extension of the Squire transformation
to Squire modes, and Ref. \cite{oberlack_new_2026} for a discussion of the Squire transformation
in terms of Lie group theory and its extension to spatiotemporal modes (modes with complex wave numbers).

\subsection{Parity symmetry \label{appendix symmetry transformation}}

For symmetric base flows $V(y)=V(-y)$, the Orr--Sommerfeld--Squire
system for a conventional fluid is parity-symmetric, i.e.,
it is symmetric under the transformation

\begin{align}
y & \rightarrow-y,\nonumber \\
u_{y} & \rightarrow-u_{y},\label{eq:symmetry transformation}\\
\varpi_{y} & \rightarrow\varpi_{y}.\nonumber 
\end{align}

Thus, the eigenmodes can be separated into two families according
to their parity: the first family has $u_{y}(y)$ that is even in $y$ and
$\varpi_{y}(y)$ that is odd in $y$, and the second family has odd $u_{y}(y)$
and even  $\varpi_{y}(y)$.

In the presence of a finite odd viscosity, the parity symmetry of the equations
is broken. However, there remains an extended symmetry that
keeps Eq. (\ref{eq:OS-SQ odd matrix operator}) invariant,
given by Eq. (\ref{eq:symmetry transformation})
together with ${\rm Ro}\rightarrow-{\rm Ro}$.
Consequently, the spectrum for symmetric
base flows does not depend on the sign of odd viscosity;
for every eigenmode at a given odd Reynolds number ${\rm Ro}$,
there is a corresponding parity-flipped one at $\left( -{\rm Ro} \right)$
[the transformation in Eq. (\ref{eq:symmetry transformation})
mapping between the two eigenmodes].

\section{Energy-conserving perturbations to non-normal operators
and transient growth \label{appendix Non-normal operators}}
In this Appendix we expand on Section \ref{instability mathematics},
defining the inner product that corresponds to the energy norm.
We show how it naturally arises from the generalized eigenvalue equation
defining the odd Orr--Sommerfeld--Squire problem,
Eq. (\ref{eq:OS-SQ odd matrix operator}) of the main text. We also
elaborate on the numerical range, and discuss the utility
of its calculation with a wider class of inner products for studying
the effects of perturbations on stability.




We follow the ideas and notations of the seminal paper on transient
growth by Reddy et al. \citep{reddy_pseudospectra_1993}. The perturbation
is represented by $\phi(t)=\left(u_{y}(t),\varpi_y(t)\right)^{T}$, where
both $u_{y},\varpi_y$ are functions of $y\in\left[-1,1\right]$. The
evolution of $\phi(t)$ is given by Eq. 
(\ref{eq:OS-SQ odd matrix operator}) in the main text, with $i \omega \rightarrow - \partial/ \partial t$,

\begin{equation}
{\cal B} \frac{\partial}{\partial t} \phi=-i {\cal A}\phi,
\label{eq:generalized time evolution}
\end{equation}
where 
\begin{align}
{\cal B} & =\begin{pmatrix}-\Delta_{k} & 0\\
0 & 1
\end{pmatrix},\\
{\cal A} & =\begin{pmatrix}-\left(\alpha V-\frac{1}{{\rm iRe}}\Delta_{k}\right)\Delta_{k}+\alpha V^{\prime\prime} & -\frac{\beta}{{\rm Ro}}\Delta_{k}\\
\beta\left(V^{\prime}-\frac{1}{{\rm Ro}}\Delta_{k}\right) & \alpha V-\frac{1}{{\rm iRe}}\Delta_{k}
\end{pmatrix}.\label{eq:cal A operator}
\end{align}

Note that the operator $\cal{B}$ is Hermitian and positive-definite
(in the usual L2 space, recalling the Dirichlet boundary conditions relevant
for the Laplacian term).
This suggests defining the following inner product,
\begin{equation}
{\left\langle \phi\vert \psi\right\rangle}_{E} \equiv \left\langle \phi\vert{\cal B}\psi\right\rangle ,
\label{eq:B inner product}
\end{equation}
where $\left\langle \cdot \vert \cdot \right\rangle $ is the standard dot inner
product. This inner product defines a norm
which turns out to be proportional to the energy of the perturbation,

\begin{align}
\left\Vert \phi\right\Vert _{E}^{2} & =\left\langle \phi\vert\phi\right\rangle _{E}=\left\langle \phi\vert{\cal B}\phi\right\rangle =\intop_{-1}^{1}\phi^{\ast}(y)\hat{{\cal B}}\phi(y){\rm d}y=\intop_{-1}^{1}\left[-u_{y}^{\ast}(y)\Delta_{k}u_{y}(y)+\varpi_y^{\ast}(y)\varpi_y(y)\right]{\rm d}y\\
& =\intop_{-1}^{1}\left[\left|\frac{\partial u_{y}(y)}{\partial y}\right|^{2}+k^{2}\left|u_{y}(y)\right|^{2}+\left|\varpi_y(y)\right|^{2}\right]{\rm d}y=k^{2}\intop_{-1}^{1}\left[ \left|u_{x}(y)\right|^2+\left|u_{y}(y)\right|^2+\left|u_{z}(y)\right|^2\right]{\rm d}y.
\label{eq:energy-norm}
\end{align}
Here, in the second line we integrated by parts (utilizing the boundary
conditions) and used the relations between $u_{x},u_{z}$ and $\partial u_{y}/\partial y,\varpi_y$
{[}Eqs. (\ref{eq:du_y_dy with u_x u_z}), (\ref{eq:eta with u_x u_z}){]}.
We thus find that $\left\Vert \phi\right\Vert _{E}^{2}$ is proportional
to the total kinetic energy of the perturbation.
Let us also write Eq. (\ref{eq:generalized time evolution}) in the standard form (note that the strict
positivity of the operator $\mathcal{B}$ 
assures its invertibility),

\begin{equation}
\frac{\partial}{\partial t} \phi=-i {\cal B}^{-1} {\cal A}\phi \equiv -i {\cal S}\phi.
\label{eq:generalized evolution to simple evolution}
\end{equation}

Note that Hermiticity under the energy norm for ${\cal S}$ is equivalent
to Hermiticity under the L2 norm for ${\cal A}$. For, if ${\cal S}^{\dagger_{E}}={\cal S}$,
then for any two states $\phi$ and $\psi$ we have

\begin{equation}
\left\langle {\cal S}\phi\vert\psi\right\rangle _{E}=\left\langle \phi\vert{\cal S}\psi\right\rangle _{E}\Longleftrightarrow\left\langle {\cal S}\phi\vert{\cal B}\psi\right\rangle =\left\langle \phi\vert{\cal B}{\cal S}\psi\right\rangle \Longleftrightarrow\left\langle {\cal B}^{-1}{\cal A}\phi\vert{\cal B}\psi\right\rangle =\left\langle \phi\vert{\cal B}{\cal B}^{-1}{\cal A}\psi\right\rangle ,
\end{equation}
and from the Hermiticity of ${\cal B}$, we get that the chain of
equalities is equivalent to $\left\langle {\cal A}\phi\vert\psi\right\rangle =\left\langle \phi\vert{\cal A}\psi\right\rangle $,
and therefore finally ${\cal S}^{\dagger_{E}}={\cal S} \Longleftrightarrow {\cal A}={\cal A}^{\dagger}$. 

Note that Hermitian conjugation in the standard inner product corresponds to the
usual operation of taking conjugate transpose. It is straightforward to see
that the odd-viscosity part of ${\cal A}$ [the part proportional
to $1/\rm{Ro}$ in Eq. (\ref{eq:cal A operator})] is Hermitian in the standard inner product,
and thus its corresponding part in $\cal S$ is Hermitian in the energy-norm inner product.
This is an immediate way to confirm the energy-conserving nature of odd viscosity.

The definition of the numerical range introduced in the main text
[Eq. (\ref{range})] is evidently
dependent on the definition of the inner product. 
As mentioned in the main text, the energy-norm inner product was utilized 
in order to utilize the energy-conservation of odd viscosity (calculating the numerical range of the full evolution operator $\mathcal S$).
In the spirit of the above transitions between the operators
$\mathcal{A}$ and $\mathcal S$ via Eq. (\ref{eq:generalized evolution to simple evolution}),
the range $F_{E}(\mathcal S)$ (subscript $E$ denoting that we are utilizing the energy-norm inner product) is equivalent to


\begin{equation}
F_{E}(\mathcal S)=
\left\{
 \frac {\left\langle \phi\vert \mathcal S \phi\right\rangle _{E}}{ \left\langle \phi\vert \phi\right\rangle _{E}} : \phi \neq 0
\right\}
=
\left\{
\frac {\left\langle \phi\vert \cal{A} \phi\right\rangle}{ \left\langle \phi\vert \phi\right\rangle _{E}} :  \phi \neq 0
\right\}.
\label{eq:energy range transformation}
\end{equation}
Numerically, calculating the right-hand side is preferable in order to avoid matrix inversions.
It is also useful for implementing the classic algorithm for the calculation of
the numerical range \cite{johnson_numerical_1978},
where one needs to calculate the Hermitian part of $\mathcal{A}_{\theta}=\exp\left( i \theta \right) \mathcal{A}$ in order to construct
tangent lines to the numerical range that are perpendicular to the ray
$\arg(z)=-\theta$ in the complex plane (see Appendix \ref{Appendix subsec range algorithm}).

In the main text, we have utilized the definition of the numerical range
to determine the potential for growth in a way that is invariant to the
odd viscosity perturbation. This suggests a general strategy for assessing the
potential relevance of perturbations in dynamical systems.
One should first compute the numerical range of the unperturbed
evolution operator using the inner product in which
the perturbation is Hermitian, and determine
whether the range protrudes
into the unstable half-plane.
If it does, the perturbation is potentially dangerous and may induce an instability
by promoting a transiently growing disturbance into an unstable eigenmode.
If it does not, the perturbation cannot generate an instability.


This discussion provides a translation of the classic approach
to stability, based on conserved quantities and Lyapunov functions,
into the language of pseudospectra and non-normal operators.
Our framework thus establishes a connection between the sensitivity
of non-normal operators to perturbations
and the classic theory of Lyapunov stability.

\section{Path-dependent adiabatic evolution and exceptional points}
\label{Exceptional pts appendix}

Here we discuss the path dependence of the adiabatic evolution of the spectrum
and its connection to exceptional points in the spectrum of the OSS operator.


First, let us briefly explain the concept of adiabatic evolution,
using the present context of the OSS system as an example.
Adiabatic evolution refers to the continuous change in the spectrum as the parameters
of an operator change continuously. Considering the OSS equation [Eq. (\ref{eq:OS-SQ odd matrix operator}) of the main text], 
let us vary two parameters of the equation with the other parameters being fixed.
For concreteness, we take ${\rm Ro}$ and $\alpha$ to be the varying parameters,
denoting $\boldsymbol{x}=\left({\rm Ro},\alpha\right)$.
When $\boldsymbol{x}$ is changed continuously, the spectrum $\left\{ c_{n}\right\}$
and eigenmodes $\left\{ \phi_{n}(y) = \left(u_{n}(y),\varpi_{n}(y)\right)^{T}\right\}$
change continuously as well \cite{kato_perturbation_1966}. This is just the statement that if the operator is well-behaved,
for a small variation $\Delta \boldsymbol{x}$ in some direction in parameter space,
the change in the spectrum goes to zero as the magnitude of $\Delta \boldsymbol{x}$ goes to zero.
This allows one to define a continuous map describing the evolution of the spectrum $\left\{ c_{n}(t)\right\}$ (and similarly, the eigenvectors) along the path
$\boldsymbol{x}(t)$, where $t$ parametrizes this path. For Hermitian operators,
this mapping is guaranteed to be well-defined,
since the eigenvectors are always orthogonal,
maintaining their unique identification along the path.

For non-Hermitian operators, an obstruction to defining such a continuous map may arise
in the case where the adiabatic path crosses an exceptional point (EP) \cite{berry_physics_2004, heiss_exceptional_2004, xue_essay_2026}.
EPs are points where two or more eigenvalues coalesce
simultaneously with their respective eigenvectors becoming exactly parallel.
If the parameter path crosses an EP,
the adiabatic mapping is no longer uniquely defined for
the path segment that is past the EP.

While having the system at the immediate vicinity of an isolated EP requires fine-tuning,
the topological consequences of an EP can be detected without directly observing the degeneracy.
For, when an adiabatic trajectory $\boldsymbol{x}$
follows a closed trajectory that encircles an EP,
the individual eigenvalues $\left\{c_{n}\right\}$
may not all return to their original values, but
instead undergo eigenvalue switching. Thus, the adiabatic evolution along a closed loop
around an EP corresponds to a map
from the eigenvalues to themselves
(which in the absence of EPs,
reduces to the identity map).
An example of an evolution along
such trajectory is presented in Fig.~\ref{fig:Eigenvalue switching}.
The eigenvalue switching pattern is rich,
with some eigenvalues simply switching between themselves,
while other eigenvalues undergo more complex permutations.
We find similar behavior occurring quite generally in the parameter space.

As a consequence, for a generic evolution of the spectrum, one cannot uniquely
establish a correspondence between eigenmodes (by means of adiabatic evolution)
at two different points of the parameter space.
As an example, consider the marginal stability surfaces at the different values of $\rm{Ro}$
in Fig.~\ref{fig:growth rate alpha-beta plane} of the main text (black curves in the figure).
A marginal mode along one curve (e.g., at the onset of the odd viscous instability, at $\rm{Ro}=10^6$)
cannot be uniquely identified
with a corresponding mode along a second marginal curve (e.g., where the 
destabilizing effect of odd viscosity is near its maximum, at $\rm{Ro}=400$).

In the context of hydrodynamics, EPs arise already for the conventional 2D Orr-Sommerfeld operator
[the limit $\beta=1/\rm{Ro}=0$ in the generalized OSS equation (\ref{eq:OS-SQ odd matrix operator}) of the main text]. 
Early works focused on resonant growth due to the degenerate eigenmodes which result from the
mode coalescence at the EPs \cite{gustavsson_excitation_1986,shanthini_degeneracies_1989}.
However, this resonance was later understood to be of lesser importance
for transient growth in channel flows \cite{reddy_energy_1993}.
Recently, subharmonic orbits surrounding EPs have been studied for
pulsating Poiseuille flow \citep{kern_subharmonic_2022}. There,
the closed period in the parameter space of the operator arises
naturally from the time-periodicity of the base flow. The parameter that
is varied for that system is the base flow
$V(y)$, which goes through a closed trajectory in the space of functions.
In our case, the closed trajectory takes place in a simpler space -- the 2D plane
representing the two scalar parameters that are being varied.

\begin{figure}[H]
\begin{centering}
\includegraphics[scale=0.5]{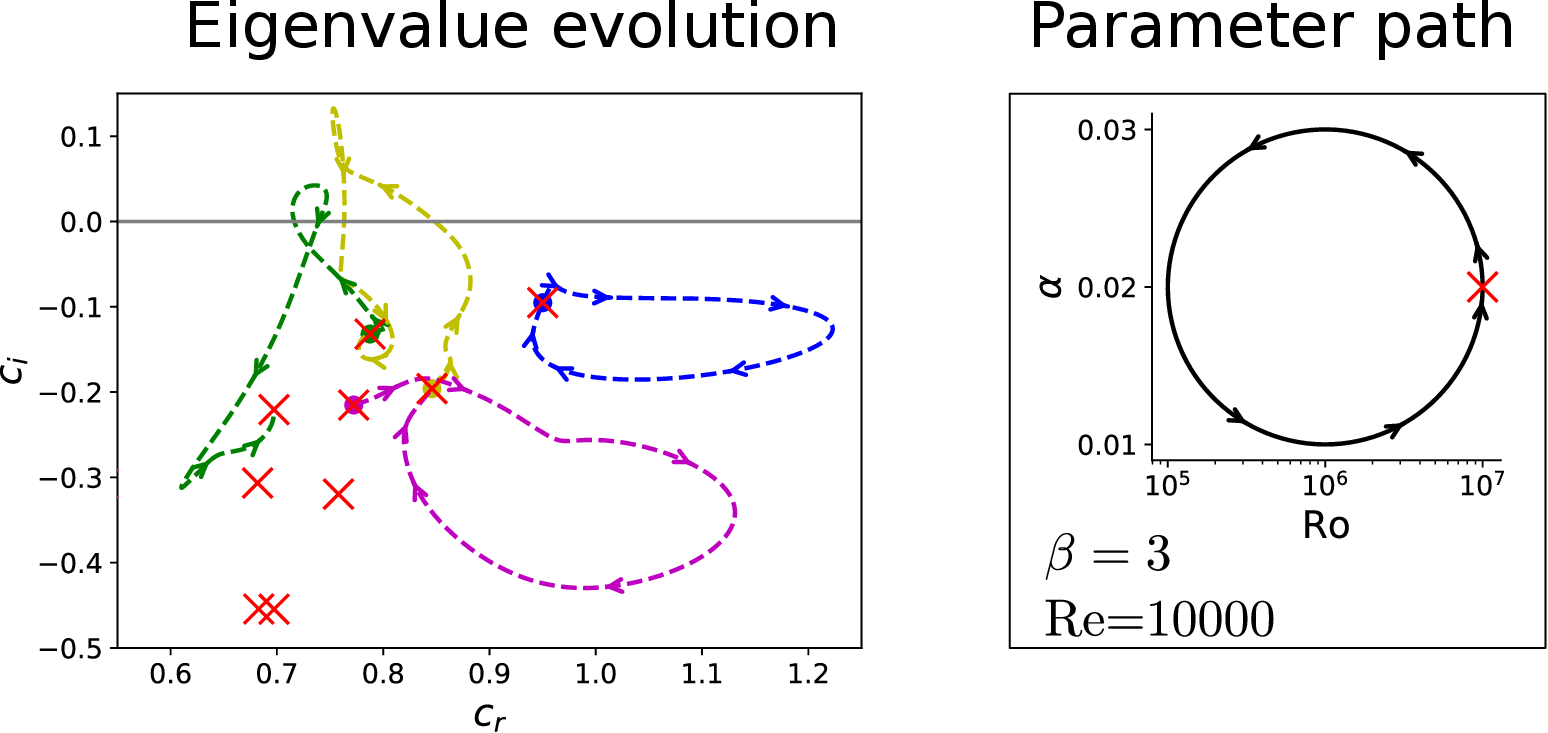}
\par\end{centering}
\caption{Continuous evolution of eigenvalues for a closed trajectory in parameter
space [circular trajectory in the (rescaled) $\alpha - \log \rm{Ro}$ plane, centered
around the point $\left( \alpha, \rm{Ro} \right) = \left( 0.02, 10^6 \right)$].
Non-closed trajectories in the eigenvalue evolution indicate eigenvalue
switching.
\label{fig:Eigenvalue switching}}
\end{figure}

\section{Analytic limits of spanwise perturbations \label{appendix analytics}}

Here, we derive some analytic limits of the equation for spanwise
perturbations, Eqs. (\ref{eq:spanwise limit}) and (\ref{eq:spanwise limit BC})
of the main text. Assuming the validity of the principle of exchange of stabilities
allows one to make analytic progress, as substituting $\omega=0$ leads to

\begin{align}
\Delta_{k}\left[\Delta_{k}^{2}+\frac{\left(\beta{\rm Re}\right)^{2}}{{\rm Ro}}\left(V^{\prime}-\frac{1}{{\rm Ro}}\Delta_{k}\right)\right]u_{y} & =0,\label{eq:spanwise limit omega zero}\\
u_{y}\left(\pm1\right)=\left[\frac{\partial}{\partial y}u_{y}\right]_{y=\pm1}=\left[\left(\Delta_{k}-\left(\frac{\beta{\rm Re}}{{\rm Ro}}\right)^{2}\right)\Delta_{k}u_{y}\right]_{y=\pm1} & =0.\label{eq:spanwise limit omega zero BC}
\end{align}
Denoting the square brackets in Eq. (\ref{eq:spanwise limit omega zero})
by

\begin{equation}
M\equiv\Delta_{k}^{2}+\frac{\left(\beta{\rm Re}\right)^{2}}{{\rm Ro}}\left(V^{\prime}-\frac{1}{{\rm Ro}}\Delta_{k}\right),
\end{equation}
we note that the solutions of Eq. (\ref{eq:spanwise limit omega zero})
are given by $Mu_{y}=0$ (4 solutions) and $Mu_{y}=\exp\left(\pm\beta y\right)$
(so that $\Delta_{k}Mu_{y}=\left(\partial_{y}^{2}-\beta^{2}\right)Mu_{y}=0$).
The last two solutions are incompatible with the boundary conditions
in Eq. (\ref{eq:spanwise limit omega zero BC}), which demand $Mu_{y}=0$
at the boundaries. Therefore, the problem reduces to the 4th-order
boundary-value problem,

\begin{align}
\left(\Delta_{k}^{2}+\frac{\left(\beta{\rm Re}\right)^{2}}{{\rm Ro}}\left(V^{\prime}-\frac{1}{{\rm Ro}}\Delta_{k}\right)\right)u_{y} & =0,\label{eq:spanwise 4th order}\\
u_{y}\left(\pm1\right)=\left[\frac{\partial}{\partial y}u_{y}\right]_{y=\pm1} & =0.
\end{align}

It can be seen that solutions of the reduced equation automatically
satisfy the two extra boundary conditions in Eq. (\ref{eq:spanwise limit omega zero BC}).

\subsection*{The inviscid limit ${\rm Re}\rightarrow\infty$ and restabilization by odd viscosity}

By taking the limit ${\rm Re\rightarrow\infty}$ at finite $\beta$
and ${\rm Ro}$, Eq. (\ref{eq:spanwise 4th order}) becomes a boundary-layer
equation, with the small parameter,

\begin{equation}
\epsilon=\frac{\left|{\rm Ro}\right|}{\beta{\rm Re}},
\end{equation}
defining the width of the boundary layer. The equation then has 4
solutions; borrowing the terminology used for the solutions of the
(conventional) Orr-Sommerfeld equation \citep{drazin_hydrodynamic_2004},
there are two ``viscous'' solutions with $\phi(y)\simeq\exp\left(\pm y/\epsilon\right)$
which describe the formation of thin boundary layers, and two ``inviscid''
solutions which solve the second-order differential equation

\begin{align}
\left(\Delta_{k}-{\rm Ro}V^{\prime}\right)u_{y} & =0,\label{eq:spanwise Re-infinity case}\\
u_{y}(\pm1) & =0.\label{eq:spanwise Re-infinity BC}
\end{align}

Let us denote the inviscid solutions by $\phi_{1,2}$. In order
to account for the two extra boundary conditions on the derivative,
these solutions should be supplemented by an $O(\epsilon)$ correction
of the viscous solutions $\phi_{3,4}$; this is possible because

\begin{equation}
\left \vert \frac{\partial\phi_{3,4}}{\partial y} \right \vert \simeq\frac{1}{\epsilon}\phi_{3,4},
\end{equation}
so a small $O(\epsilon)$ correction of $\phi_{3,4}$ gives an
$O(1)$ correction to the derivative.

\subsubsection*{Solutions and critical ${\rm Ro}$ for Poiseuille flow}

For the Poiseuille flow, $V^{\prime}=-2y$, and Eq. (\ref{eq:spanwise Re-infinity case})
reduces to the Airy equation plus an additional term,

\begin{equation}
\left(\frac{\partial^{2}}{\partial y^2}-\beta^{2}+2{\rm Ro}y\right)\phi(y)=0
\end{equation}
Rescaling $z=\lambda y$, $\tilde{\phi}(z)=\phi(y)$ with $\lambda=-\left(2{\rm Ro}\right)^{1/3}$,
we obtain

\begin{equation}
\left(\frac{\partial^{2}}{\partial z^2}-z-\frac{\beta^{2}}{\lambda^2}\right)\tilde{\phi}(z)=0.
\end{equation}

In the long-wavelength limit $\beta\rightarrow0$, the solutions are given by the
Airy functions,

\begin{equation}
\tilde{\phi}(z)=A{\rm Ai}(z)+B{\rm Bi}(z).
\end{equation}
The boundary conditions {[}Eq. (\ref{eq:spanwise Re-infinity BC}){]}
then give the condition

\begin{equation}
{\rm Ai}(\lambda){\rm {\rm Bi}}(-\lambda)-{\rm Ai}(-\lambda){\rm {\rm Bi}}(\lambda)=0.
\end{equation}
The first non-trivial solution is given by

\begin{equation}
\lambda\approx-2.341,
\end{equation}
corresponding to

\begin{equation}
{\rm Ro}_{{\rm c}}=-\frac{1}{2}\lambda^{3}\approx6.412.
\end{equation}

We have found a minimal value of ${\rm Ro}$ for which there is a
marginal stationary solution ($\omega=0$) for spanwise modes in the inviscid limit ${\rm Re\rightarrow\infty}$.
Therefore (assuming the principle of exchange of stabilities),
when increasing the odd viscosity further such that ${\rm Ro}<{\rm Ro}_{{\rm c}}$,
the odd viscous instability disappears.
This is the restabilization threshold that we obtain also
in our numerical results (the vertical asymptote of the
black curve in Fig. \ref{fig:Critical curves Re(Ro)}
of the main text).

\section{Numerical methods}

In this Appendix, we provide details on the pseudospectral
method used to solve the generalized eigenvalue problem and to
calculate the numerical range of the Orr-Sommerfeld-Squire operator.
While the pseudospectral method is standard and well documented in
the context of the Orr--Sommerfeld equation \citep{boyd_chebyshev_2013},
we write this Appendix in a relatively self-contained manner to address
two subtle technical points relevant for this work:

1. The different
treatment of the numerical approximations of the wall-normal velocity
and vorticity, $u(y)$ and $\varpi(y)$ (using the modification due
to Huang and Sloan \citep{huang_pseudospectral_1994} for the first
function, but not the second).

2. The importance (for the calculation
of the numerical range via the pseudospectral method) of
using an approximating subspace which is invariant under
the derivative operator. In other words, we will show that it is important
that taking the derivative of any approximating basis function results
in a linear combination of the other basis functions.

\subsection{Pseudospectral collocation method}
We use a pseudospectral method with the modification due to Huang
and Sloan \citep{huang_pseudospectral_1994,weideman_matlab_2000},
which we briefly review. We are interested in numerically solving
the differential equation {[}Eq. (\ref{eq:OS-SQ odd matrix operator}) of
the main paper{]}, which is a generalized eigenvalue problem on the
domain $[-1,1]$. For simplicity, we first describe the solution of
the simpler, conventional 2D Orr--Sommerfeld problem,
detailing the additional steps for the full
3D equation in the next section. Let us write the generalized eigenvalue
problem as {[}In this Appendix, we denote the spatial variable by
$x$ ($y$ in the main text){]}

\begin{equation}
\mathcal{A}u(x)=\lambda \mathcal{B}u(x),\label{eq:eigval equation numerics}
\end{equation}
with $\mathcal A$ and $\mathcal B$ being differential operators of fourth and second
order, respectively, and $\lambda$ being the eigenvalue. In the pseudospectral
method, the continuous domain is approximated by $N$ collocation
points,

\begin{equation}
-1=x_{1}<x_{2}<\ldots<x_{N-1}<x_{N}=1.
\end{equation}
Define the cardinal functions $\ell_{k}(x)$ as $\left(N-1\right)$-degree
polynomials satisfying $\ell_{k}(x_{j})=\delta_{kj},\quad1\le j,k\le N$.
Here, we choose the collocation points as Chebyshev--Gauss--Lobatto
nodes, so that $x_{1},\ldots,x_{N}$ correspond to the extremal points
on $\left[-1,1\right]$ of the Chebyshev polynomial of the first kind
$T_{N-1}$. In this case, the cardinal functions are given by \citep{weideman_matlab_2000}

\begin{equation}
\ell_{k}(x)=\frac{\left(-1\right)^{N-k+1}}{c_{k}}\frac{1-x^{2}}{\left(N-1\right)^{2}}\frac{T_{N-1}^{\prime}(x)}{x-x_{k}},
\end{equation}
with $c_{1}=c_{N}=2$ and $c_{k}=1$ for $2\le k\le N-1$. In the
Orr--Sommerfeld equation, the boundary conditions for the wall-normal
velocity function {[}denoted in this Appendix by $u(x)${]} involve
four equations,

\begin{equation}
u(x_{1})=u(x_{N})=u'(x_{1})=u'(x_{N})=0.\label{eq:u BC}
\end{equation}

The Dirichlet boundary conditions on $u(x)$ constrain the solutions
to have zero coefficients for the cardinal functions of the end points,
$\ell_{1}(x)$ and $\ell_{N}(x)$. To account for the boundary conditions
on $u'(x)$, Huang and Sloan introduced the modified basis functions
\citep{huang_pseudospectral_1994},

\begin{equation}
h_{k}(x)=\frac{\left(x-x_{1}\right)\left(x-x_{N}\right)}{\left(x_{k}-x_{1}\right)\left(x_{k}-x_{N}\right)}\ell_{k}(x),\qquad2\le k\le N-1.\label{eq:Huang-sloan base functions}
\end{equation}

To eliminate the appearance of spurious eigenvalues, Ref. \citep{huang_pseudospectral_1994}
uses two different approximations for the functions $u(x)$ on the
two sides of Eq. (\ref{eq:eigval equation numerics}): on the left-hand
side, the function is expanded with the basis $\left\{ h_{k}\right\} $,
satisfying the full boundary conditions,

\begin{equation}
u^{N}(x)=\sum_{i=2}^{N-1}u_{i}h_{i}(x),\label{eq:Huang-Sloan expansion}
\end{equation}
while for the right-hand side, the function is expanded with $\left\{ \ell_{k}\right\} $,
satisfying just the two homogeneous Dirichlet boundary conditions,

\begin{equation}
\tilde{u}^{N}(x)=\sum_{i=2}^{N-1}u_{i}\ell_{i}(x),
\end{equation}
where $u_{i}=u(x_{i})$. The matrix representations of the differential
operators for the appropriate interpolants are given by

\begin{align}
\text{\ensuremath{D_{ij}}} & \equiv\frac{d\ell_{j}}{dx}(x_{i}), \qquad 1\le i,j \le N,\\
\tilde{D}_{ij} & \equiv\frac{dh_{j}}{dx}(x_{i})=\frac{2x_{j}-x_{1}-x_{N}}{\left(x_{j}-x_{1}\right)\left(x_{j}-x_{N}\right)}\delta_{ij}+\frac{\left(x_{i}-x_{1}\right)\left(x_{i}-x_{N}\right)}{\left(x_{j}-x_{1}\right)\left(x_{j}-x_{N}\right)}\text{\ensuremath{D_{ij}}}, \qquad 1\le i \le N, 2\le j \le N-1,\label{eq:D tilde derivative}
\end{align}
with the first and last columns ($j=1,N$) of $\tilde D _ {ij}$ being irrelevant
(these are to be truncated, as explained in the following).
For the higher-order derivatives of the functions $\ell_{k}(x)$,
one may raise the matrix $D$ to higher powers; for the higher-order
derivatives of $h_{k}(x)$, we calculate recurrence relations with
the matrix $D_{ij}$ as in Eq. (\ref{eq:D tilde derivative}), differentiating
Eq. (\ref{eq:Huang-sloan base functions}) using the product rule (raising
$\tilde{D}$ to higher powers is erroneous since the derivative takes
functions out of the space spanned by $\left\{ h_{k}(x) \right\}$; see also the later
section on the calculation of the numerical range). Then, one calculates
a matrix representation for the operators $\mathcal A$ and $\mathcal B$, representing
the derivatives in each operator by $\tilde{D}$ and $D$, respectively.
To impose the boundary conditions, the first and last rows and
columns of the matrices ($\mathcal A_{1j},\mathcal A_{Nj},\mathcal A_{i1},\mathcal A_{iN}$ and similarly
for $\mathcal B$) are truncated to form $\left(N-2\right)\times\left(N-2\right)$
matrices, resulting in a generalized matrix eigenvalue problem corresponding
to Eq. (\ref{eq:eigval equation numerics}). The eigenvalue problem
is then solved numerically. For explicit formulas and the MATLAB implementation,
see Ref. \citep{weideman_matlab_2000}, with the solution to the Orr--Sommerfeld
equation using the Huang--Sloan modification given in Section 5.5
\footnote{Note that Ref. \citep{weideman_matlab_2000} uses the representation
	$\tilde{D}$ only for the 4th-order derivative in Eq. (\ref{eq:eigval equation numerics}),
	while we follow Ref. \citep{huang_pseudospectral_1994} and use $\tilde{D}$
	for all the derivative terms on the left-hand side of the equation.}.

\subsubsection*{Application to the 3D Orr-Sommerfeld-Squire problem}

For the Orr--Sommerfeld--Squire equation solved in this work, the
generalized eigenvalue problem can be written as

\begin{equation}
\mathcal A\begin{pmatrix}u(x)\\
\varpi(x)
\end{pmatrix}=\lambda \mathcal B\begin{pmatrix}u(x)\\
\varpi(x)
\end{pmatrix},\label{eq:generalized eigval numerics}
\end{equation}
where $\mathcal A$ and $\mathcal B$ are $2\times2$ matrices of differential operators,
with $\mathcal B$ being block diagonal,

\begin{equation}
\mathcal B=\begin{pmatrix}-\Delta_{k} & 0\\
0 & 1
\end{pmatrix}.\label{eq:operator B}
\end{equation}
The boundary conditions on $u(x)$ are the same as in the previous
part {[}Eq. (\ref{eq:u BC}){]}, while $\varpi(x)$ satisfies the
two homogeneous Dirichlet boundary conditions $\varpi(x_{1})=\varpi(x_{N})=0$.
For $u(x)$ we use two different approximations on the two sides of
the equation (\ref{eq:generalized eigval numerics}) as in the 2D
Orr--Sommerfeld case, while for $\varpi(x)$ we use the same approximation
on both sides,

\begin{equation}
\varpi(x)\approx\varpi^{N}(x)=\sum_{i=2}^{N-1}\varpi_{i}\ell_{i}(x).
\end{equation}

The different treatment for $u$ and $\varpi$ is because it is the
Laplacian part of the operator $\mathcal B$ {[}first row in Eq. (\ref{eq:operator B}){]}
that leads to spurious eigenvalue issues; if one uses the modified
basis {[}Eq. (\ref{eq:Huang-Sloan expansion}){]} in the approximation
of $u(x)$ on the right-hand side of the generalized eigenvalue equation,
after performing truncation, the relevant part of $\mathcal B$ (top-left block)
becomes nearly singular, leading to spurious eigenvalues \citep{huang_pseudospectral_1994};
for $\varpi(x)$, the corresponding block of $\mathcal B$ that acts on it
is just the identity matrix, and no similar problem arises.

From here we proceed as in the previous section, taking the derivative
operators in $\mathcal A$ to be the ones corresponding to the basis functions
of $u^{N}(x)$ and $\varpi^{N}(x)$, the ones in $\mathcal B$ to correspond
to the approximations $\tilde{u}^{N}(x)$ and $\varpi^{N}(x)$, and
truncating the first and last rows and columns in each of the four $N\times N$
blocks of the two matrices, giving a matrix equation of the form

\begin{equation}
\begin{pmatrix}\mathcal A_{uu} & \mathcal A_{u\varpi}\\
\mathcal A_{\varpi u} & \mathcal A_{\varpi\varpi}
\end{pmatrix}\begin{pmatrix}u\\
\varpi
\end{pmatrix}=\lambda\begin{pmatrix}\mathcal B_{uu} & 0\\
0 & \mathcal B_{\varpi\varpi}
\end{pmatrix}\begin{pmatrix}u\\
\varpi
\end{pmatrix},\label{eq:eigenvalue eq methods}
\end{equation}
where $u$ and $\varpi$ are $\left(N-2\right)$-component vectors
with components corresponding to their values at the interior nodes,
and the blocks making up $\mathcal A$ and $\mathcal B$ are of dimensions $\left(N-2\right)\times\left(N-2\right)$.

\subsection{Calculation of the numerical range}

\subsubsection*{Invariance of the representation space under the operator in the
	pseudospectral method}

Since we have calculated a matrix representation of an operator (in
the pseudospectral representation) for the purpose of solving the
generalized eigenvalue problem, it may seem natural to utilize that
representation for the calculation of the numerical range {[}i.e., the matrices
$\mathcal A$ and $\mathcal B$ of Eq. (\ref{eq:eigenvalue eq methods}){]}. However,
some care is required when utilizing the same matrices for the calculation
of the numerical range, since a naive application (projecting all operations
on the discrete basis and reinterpolating) may violate the fundamental
anti-Hermiticity property of the first-derivative operator (under
Dirichlet boundary conditions). This is because the space spanned
by $\left\{ \ell_{1}(x),h_{2}(x),\ldots,h_{N-1}(x),\ell_{N}(x)\right\} $
is not invariant under the derivative operator.

To see why this is a problem, note that the elements of the derivative
matrix $\tilde{D}_{ij}$ defined in Eq. (\ref{eq:D tilde derivative})
represent $h_{j}^{\prime}(x_{i})$, where $x_{i}$ is a Chebyshev--Gauss--Lobatto
node. However, if the derivative takes the function outside of the
space spanned by $\left\{ h_{k}(x)\right\} $ , then $\tilde{D}$
is not sufficient to reproduce the function $h_{j}^{\prime}(x)$;
i.e., we have

\begin{equation}
h_{j}^{\prime}(x)=\sum_{i=2}^{N-1}\tilde{D}_{ij}h_{i}(x)+e_{j}(x),
\end{equation}
with $e_{j}(x)\neq0$ in general. As a result, we cannot calculate matrix elements
such as $\left\langle h_{i}\vert\partial_{x}h_{j}\right\rangle $
with knowledge of only $\tilde{D}$ (and the Gram matrix $M_{ij}\equiv\left\langle h_{i}\vert h_{j}\right\rangle $).
Note that if the residual to the projection is orthogonal to the cardinal
functions, $\left\langle h_{i}\vert e_{j}\right\rangle =0$ for all
$\left(i,j\right)$, a problem does not arise; however, the pseudospectral
approximation $h_{j}^{\prime}(x)\approx\sum_{i}\tilde{D}_{ij}h_{i}(x)$
is generally not an orthogonal projection (as opposed, for example,
to the Galerkin method, where the residuals are orthogonal to the
approximating functions by construction \citep{boyd_chebyshev_2013}).

Now, let us consider the invariance (or lack thereof) under the derivative
operator of the subspaces used in the pseudospectral approximation.
First, let us note that for the basis $\mathcal G_{\ell}=\left\{ \ell_{1}(x),\ell_{2}(x),\ldots,\ell_{N-1}(x),\ell_{N}(x)\right\} $,
its spanned subspace is invariant under the derivative operator, and
the problem mentioned above does not arise; this is because ${\mathcal G}_{\ell}$
consists of $N$ linearly independent $\left(N-1\right)$-degree polynomials,
thus spanning the space ${\cal P}_{N-1}$ of polynomials of degree at most $\left(N-1\right)$.
This space is invariant under the derivative operator,
as the derivative simply lowers the degree of a polynomial.

On the other hand, the Huang--Sloan basis

\begin{equation}
{\cal \mathcal{G}}=\left\{ \ell_{1}(x),h_{2}(x),\ldots,h_{N-1}(x),\ell_{N}(x)\right\} 
\end{equation}
is \textit{not} invariant under the derivative operator. Unlike the
previous case, it is made of $N$ polynomials of degree at most $N+1$,
and thus it does not span the full space ${\cal P}_{N+1}$. Indeed,
by a dimensional argument, it can be seen that the span of ${\cal G}$
is not invariant under the derivative; by taking successive derivatives
of an $\left(N+1\right)$-degree polynomial, $\left(\partial_{x}\right)^{n}h_{k}(x),\quad n=1,\ldots N+1,$
we can obtain $\left(N+1\right)$ linearly independent polynomials,
ruling out the possibility that the derivative operator stays in the
$N-$dimensional subspace ${\rm span}({\cal G})$.

Therefore, to obtain invariance under the derivative, it is necessary
to complete ${\cal G}$ with two additional functions, so that its
span can contain all of ${\cal P}_{N+1}$. We build a completed basis
${\cal C}$ with a small modification, keeping the interior functions
$h_{2},\ldots h_{N-1}$ of ${\cal G}$ and adding four boundary cardinal
functions. First, define modified basis functions for the endpoints
analogous to Eq. (\ref{eq:Huang-sloan base functions}),

\begin{equation}
\tilde{h}_{i}(x)=\ell_{i}(x)\frac{x-x_{j}}{x_{i}-x_{j}},\qquad\left(i,j\right)=\left(1,N\right),\left(N,1\right),
\end{equation}
so that, for example, for $i=1$, the corresponding interpolating
function $\tilde{h}_{1}(x)$ satisfies the homogeneous value and derivative conditions,
$\tilde{h}_{1}(x_{N})=\tilde{h}_{1}'(x_{N})=0$.
Now, define the following four interpolants for the boundaries,

\begin{align}
g_{i}(x) & =\tilde{h}_{i}(x)\left(1-\tilde{h}_{i}'(x_{i})\left(x-x_{i}\right)\right),\qquad i=1,N,\\
\tilde{g}_{i}(x) & =\tilde{h}_{i}(x)\left(x-x_{i}\right),\qquad i=1,N.
\end{align}
Simple calculations show that these satisfy

\begin{align}
g_{i}(x_{j}) & =\delta_{ij},\qquad i=1,N,\quad j=1,\ldots,N,\\
g_{i}'(x_{j}) & =0,\qquad i=1,N,\quad j=1,N,\\
\tilde{g}_{i}(x_{j}) & =0,\qquad i=1,N,\quad j=1,\ldots,N,\\
\tilde{g}_{i}'(x_{j}) & =\delta_{ij},\qquad i=1,N,\quad j=1,N.
\end{align}
Now the full basis 
\begin{equation}
\mathcal{C}=\left\{ g_{1}(x),h_{2}(x),\ldots,h_{N-1}(x),g_{N}(x),\tilde{g}_{1}(x),\tilde{g}_{N}(x)\right\} 
\end{equation}
is sufficient to represent the $\left(N+2\right)$ degrees of freedom
$\left\{ f(x_{1}),\ldots,f(x_{N}),f'(x_{1}),f'(x_{N})\right\} $.
As a basis consisting of $\left(N+2\right)$ linearly independent
polynomials of degree $\left(N+1\right)$, $\mathcal{C}$ spans the
full space ${\cal P}_{N+1}$. Consequently, ${\rm span({\cal C})}$
is invariant under the derivative operator. We remind the reader that the differential operators of interest
include Dirichlet boundary conditions, and thus their domain is
limited to the subspace spanned by the interior functions; nonetheless, the non-invariance that we have discussed
means that the range is generally not limited to the interior subspace.
It is necessary to calculate the projections of an operator $L$ to the full space
spanned by $\mathcal C$, since calculating
inner products $\left\langle h_{i}\vert Lh_{j}\right\rangle$ involves terms like

\begin{equation}
\left\langle h_{i}\vert Lh_{j}\right\rangle =L_{g_{1},h_{j}}\left\langle h_{i}\vert g_{1}\right\rangle +\cdots,
\end{equation}
where $L_{g_{1},h_{j}}$ is the coefficient corresponding to $g_1$ in the expansion of the projection of $L h_j$ to $\rm {span} \left( \mathcal C \right)$.

To summarize, in order to ensure the basic anti-Hermitian property of
the derivative operator (for functions satisfying Dirichlet boundary
conditions) in the pseudospectral method, it is necessary to discretize the operator with a basis
whose span is invariant under it; the Huang--Sloan modified basis {[}Eq.
(\ref{eq:Huang-sloan base functions}){]} does not have this property,
and therefore it must be completed as we have done above (or alternatively,
a different basis should be used for discretization).

\subsubsection*{Algorithm for calculation of the numerical range \label{Appendix subsec range algorithm}}

For the calculation of the numerical range, we implement the algorithm by
Johnson \cite{johnson_numerical_1978}, which we briefly describe here.
For the explanation,
let us first consider the numerical range to be defined with respect
to the standard inner product, deferring the general case until later.
The algorithm utilizes the fact that the maximal real extent of the
numerical range of an operator $\mathcal A$ is identical to that of its Hermitian part
(the imaginary version of this property is a main
theme in our work),

\begin{align}
\underset{\left\Vert \phi\right\Vert =1}{{\rm sup}}{\rm Re}\left[\left\langle \phi\vert \mathcal A\phi\right\rangle \right] & =\underset{\left\Vert \phi\right\Vert =1}{{\rm sup}}{\rm Re}\left[\left\langle \phi\vert \mathcal A_{H}\phi\right\rangle \right],\label{eq:range Hermitian identity}\\
\mathcal A_{H} & \equiv\frac{\mathcal A+\mathcal A^{\dagger}}{2}.
\end{align}
Since $\mathcal A_{H}$ is Hermitian, its numerical range is the convex hull of its eigenvalues
$\left\{ \lambda_{n}\right\} $, and 
\begin{equation}
\underset{\left\Vert \phi\right\Vert =1}{{\rm sup}}\left({\rm Re}\left[\left\langle \phi\vert \mathcal A_{H}\phi\right\rangle \right]\right)=\max\left(\lambda_{n}\right)\equiv\lambda_{M}.\label{eq:eigval eq in range}
\end{equation}
Thus, the line
\begin{equation}
L_{0}=\left\{ \lambda_{M}+ti:t\in\mathbb{R}\right\} \label{eq:vertical line range}
\end{equation}
is tangent to the numerical range of $\mathcal A$. The key idea of Ref. \cite{johnson_numerical_1978}
is that similarly, one can calculate tangent lines to the numerical range 
\begin{equation}
L_{\theta}=\left\{ e^{-i\theta}\left(\lambda_{M}(\theta)+ti\right):t\in\mathbb{R}\right\} \label{eq:line theta range}
\end{equation}
by considering

\begin{equation}
\underset{\left\Vert \phi\right\Vert =1}{{\rm sup}}{\rm Re}\left[e^{i\theta}\left\langle \phi\vert \mathcal A\phi\right\rangle \right]=\underset{\left\Vert \phi\right\Vert =1}{{\rm sup}}{\rm Re}\left[\left\langle \phi\vert e^{i\theta}\mathcal A\phi\right\rangle \right]=\underset{\left\Vert \phi\right\Vert =1}{{\rm sup}}{\rm Re}\left[\left\langle \phi\vert\left(\frac{e^{i\theta}\mathcal A+e^{-i\theta}\mathcal A^{\dagger}}{2}\right)\phi\right\rangle \right],
\end{equation}
and calculating the largest eigenvalue $\lambda_{M}(\theta)$ of the
Hermitian matrix
\begin{equation}
\mathcal A_{H}(\theta)\equiv\frac{e^{i\theta}\mathcal A+e^{-i\theta}\mathcal A^{\dagger}}{2}.
\end{equation}

Thus, by calculating the largest eigenvalues $\lambda_{M}(\theta)$
of $\mathcal A_{H}(\theta)$ for $\theta \in \left[ 0, 2\pi \right)$,
one can reconstruct the numerical range of $\mathcal A$ \cite{johnson_numerical_1978}.
For the numerical
calculation producing Fig. \ref{fig:mechanism and range} of the main text,
we use a finite grid $\left\{ \theta_{n}\right\} $
with $N_{\theta}=350$ points.

Now, let us discuss the generalization of the algorithm to a general
inner product. In Appendix \ref{appendix Non-normal operators}, we have shown that the numerical range of an
operator $\mathcal S$ with respect to the energy-norm inner product,
\begin{equation}
\left\langle \varphi\vert\psi\right\rangle _{E}=\left\langle \varphi\vert \mathcal B \psi\right\rangle, \label{eq:energy-norm inner product range}
\end{equation}
is equivalent to

\begin{equation}
F_{E}(\mathcal S)=\left\{ \frac{\left\langle \phi\vert \mathcal A \phi\right\rangle }{\left\langle \phi\vert\phi\right\rangle _{E}}:\phi\neq0\right\} ,
\end{equation}
where $\mathcal A=\mathcal B \mathcal S$ (note that by writing the elements of
the numerical range as a quotient, they are independent of $\left\Vert \phi\right\Vert $).
A similar transformation holds for any general inner product, related
to the standard inner product by an equation analogous to (\ref{eq:energy-norm inner product range}).
It is straightforward to see that in order to implement the algorithm
by Johnson, the step in Eq. (\ref{eq:range Hermitian identity}) needs
to be modified; we are now maximizing

\begin{equation}
\underset{\phi \neq 0}{{\rm sup}} \left ( {\rm Re}\left[\frac{\left\langle \phi\vert \mathcal A\phi\right\rangle }{\left\langle \phi\vert\phi\right\rangle _{E}}\right] \right)
=
\underset{\phi \neq 0}{{\rm sup}} \left( {\rm Re}\left[\frac{\left\langle \phi\vert \mathcal A_{H}\phi\right\rangle }{\left\langle \phi\vert\phi\right\rangle _{E}}\right] \right) ,\label{eq:generalized eigval in range}
\end{equation}
where $\mathcal A_{H}$ is the Hermitian part of $\mathcal A$ with respect
to the standard inner product. Note the relation between the expression on the right-hand side of
Eq. (\ref{eq:generalized eigval in range}) and the generalized eigenvalue
problem

\begin{equation}
\mathcal A_{H}\psi(y)=\lambda \mathcal B\psi(y),\label{eq:generalized eigval explicit range}
\end{equation}
where $\left( \lambda, \psi(y) \right)$ is an eigenvalue-eigenfunction pair;
as in the case of a regular eigenvalue problem, the quotient in the
right-hand side of Eq. (\ref{eq:generalized eigval in range}) is
maximized when $\phi$ is the eigenfunction with the largest eigenvalue
in the solutions of Eq. (\ref{eq:generalized eigval explicit range}).
Thus, for the calculation of the numerical range defined with respect
to the energy inner product, the only modification to the algorithm
described above is to take $\lambda_{M}$ for the vertical tangent
line {[}Eq. (\ref{eq:vertical line range}){]} to be the largest eigenvalue
in the generalized eigenvalue problem {[}Eq. (\ref{eq:generalized eigval explicit range}){]},
and similarly for the matrices $\mathcal A_{H}(\theta)$ that determine the
tangent lines $L_{\theta}$ {[}Eq. (\ref{eq:line theta range}){]}.

\end{widetext}

\bibliography{odd_instability_refs}

\begin{thebibliography}{66}%
\makeatletter
\providecommand \@ifxundefined [1]{%
 \@ifx{#1\undefined}
}%
\providecommand \@ifnum [1]{%
 \ifnum #1\expandafter \@firstoftwo
 \else \expandafter \@secondoftwo
 \fi
}%
\providecommand \@ifx [1]{%
 \ifx #1\expandafter \@firstoftwo
 \else \expandafter \@secondoftwo
 \fi
}%
\providecommand \natexlab [1]{#1}%
\providecommand \enquote  [1]{``#1''}%
\providecommand \bibnamefont  [1]{#1}%
\providecommand \bibfnamefont [1]{#1}%
\providecommand \citenamefont [1]{#1}%
\providecommand \href@noop [0]{\@secondoftwo}%
\providecommand \href [0]{\begingroup \@sanitize@url \@href}%
\providecommand \@href[1]{\@@startlink{#1}\@@href}%
\providecommand \@@href[1]{\endgroup#1\@@endlink}%
\providecommand \@sanitize@url [0]{\catcode `\\12\catcode `\$12\catcode
  `\&12\catcode `\#12\catcode `\^12\catcode `\_12\catcode `\%12\relax}%
\providecommand \@@startlink[1]{}%
\providecommand \@@endlink[0]{}%
\providecommand \url  [0]{\begingroup\@sanitize@url \@url }%
\providecommand \@url [1]{\endgroup\@href {#1}{\urlprefix }}%
\providecommand \urlprefix  [0]{URL }%
\providecommand \Eprint [0]{\href }%
\providecommand \doibase [0]{https://doi.org/}%
\providecommand \selectlanguage [0]{\@gobble}%
\providecommand \bibinfo  [0]{\@secondoftwo}%
\providecommand \bibfield  [0]{\@secondoftwo}%
\providecommand \translation [1]{[#1]}%
\providecommand \BibitemOpen [0]{}%
\providecommand \bibitemStop [0]{}%
\providecommand \bibitemNoStop [0]{.\EOS\space}%
\providecommand \EOS [0]{\spacefactor3000\relax}%
\providecommand \BibitemShut  [1]{\csname bibitem#1\endcsname}%
\let\auto@bib@innerbib\@empty
\bibitem [{\citenamefont {Avron}(1998)}]{avron_odd_1998}%
  \BibitemOpen
  \bibfield  {author} {\bibinfo {author} {\bibfnamefont {J.~E.}\ \bibnamefont
  {Avron}},\ }\bibfield  {title} {\bibinfo {title} {Odd {Viscosity}},\ }\href
  {https://doi.org/10.1023/A:1023084404080} {\bibfield  {journal} {\bibinfo
  {journal} {Journal of Statistical Physics}\ }\textbf {\bibinfo {volume}
  {92}},\ \bibinfo {pages} {543} (\bibinfo {year} {1998})}\BibitemShut
  {NoStop}%
\bibitem [{\citenamefont {Fruchart}\ \emph {et~al.}(2023)\citenamefont
  {Fruchart}, \citenamefont {Scheibner},\ and\ \citenamefont
  {Vitelli}}]{fruchart_odd_2023}%
  \BibitemOpen
  \bibfield  {author} {\bibinfo {author} {\bibfnamefont {M.}~\bibnamefont
  {Fruchart}}, \bibinfo {author} {\bibfnamefont {C.}~\bibnamefont
  {Scheibner}},\ and\ \bibinfo {author} {\bibfnamefont {V.}~\bibnamefont
  {Vitelli}},\ }\bibfield  {title} {\bibinfo {title} {Odd {Viscosity} and {Odd}
  {Elasticity}},\ }\href
  {https://doi.org/10.1146/annurev-conmatphys-040821-125506} {\bibfield
  {journal} {\bibinfo  {journal} {Annu. Rev. Condens. Matter Phys.}\ }\textbf
  {\bibinfo {volume} {14}},\ \bibinfo {pages} {471} (\bibinfo {year}
  {2023})}\BibitemShut {NoStop}%
\bibitem [{\citenamefont {Rosenbluth}\ \emph {et~al.}(1962)\citenamefont
  {Rosenbluth}, \citenamefont {Krall},\ and\ \citenamefont
  {Rostoker}}]{rosenbluth1962nuclear}%
  \BibitemOpen
  \bibfield  {author} {\bibinfo {author} {\bibfnamefont {M.~N.}\ \bibnamefont
  {Rosenbluth}}, \bibinfo {author} {\bibfnamefont {N.~A.}\ \bibnamefont
  {Krall}},\ and\ \bibinfo {author} {\bibfnamefont {N.}~\bibnamefont
  {Rostoker}},\ }\bibfield  {title} {\bibinfo {title} {{Finite {Larmor} radius
  stabilization of ``weakly'' unstable confined plasmas, Nuclear Fusion, 1962
  Supplement}},\ }\href@noop {} {\bibfield  {journal} {\bibinfo  {journal}
  {Part}\ }\textbf {\bibinfo {volume} {1}},\ \bibinfo {pages} {143} (\bibinfo
  {year} {1962})}\BibitemShut {NoStop}%
\bibitem [{\citenamefont {Roberts}\ and\ \citenamefont
  {Taylor}(1962)}]{roberts_magnetohydrodynamic_1962}%
  \BibitemOpen
  \bibfield  {author} {\bibinfo {author} {\bibfnamefont {K.~V.}\ \bibnamefont
  {Roberts}}\ and\ \bibinfo {author} {\bibfnamefont {J.~B.}\ \bibnamefont
  {Taylor}},\ }\bibfield  {title} {\bibinfo {title} {Magnetohydrodynamic
  {Equations} for {Finite} {Larmor} {Radius}},\ }\href
  {https://doi.org/10.1103/PhysRevLett.8.197} {\bibfield  {journal} {\bibinfo
  {journal} {Phys. Rev. Lett.}\ }\textbf {\bibinfo {volume} {8}},\ \bibinfo
  {pages} {197} (\bibinfo {year} {1962})}\BibitemShut {NoStop}%
\bibitem [{\citenamefont {Avron}\ \emph {et~al.}(1995)\citenamefont {Avron},
  \citenamefont {Seiler},\ and\ \citenamefont {Zograf}}]{avron_viscosity_1995}%
  \BibitemOpen
  \bibfield  {author} {\bibinfo {author} {\bibfnamefont {J.~E.}\ \bibnamefont
  {Avron}}, \bibinfo {author} {\bibfnamefont {R.}~\bibnamefont {Seiler}},\ and\
  \bibinfo {author} {\bibfnamefont {P.~G.}\ \bibnamefont {Zograf}},\ }\bibfield
   {title} {\bibinfo {title} {Viscosity of {Quantum} {Hall} {Fluids}},\ }\href
  {https://doi.org/10.1103/PhysRevLett.75.697} {\bibfield  {journal} {\bibinfo
  {journal} {Phys. Rev. Lett.}\ }\textbf {\bibinfo {volume} {75}},\ \bibinfo
  {pages} {697} (\bibinfo {year} {1995})}\BibitemShut {NoStop}%
\bibitem [{\citenamefont {Read}(2009)}]{read_non-abelian_2009}%
  \BibitemOpen
  \bibfield  {author} {\bibinfo {author} {\bibfnamefont {N.}~\bibnamefont
  {Read}},\ }\bibfield  {title} {\bibinfo {title} {Non-{Abelian} adiabatic
  statistics and {Hall} viscosity in quantum {Hall} states and $p_x + i p_y$
  paired superfluids},\ }\href {https://doi.org/10.1103/PhysRevB.79.045308}
  {\bibfield  {journal} {\bibinfo  {journal} {Phys. Rev. B}\ }\textbf {\bibinfo
  {volume} {79}},\ \bibinfo {pages} {045308} (\bibinfo {year}
  {2009})}\BibitemShut {NoStop}%
\bibitem [{\citenamefont {Srivastava}\ and\ \citenamefont
  {Mukerjee}(2025)}]{srivastava_electron_2025}%
  \BibitemOpen
  \bibfield  {author} {\bibinfo {author} {\bibfnamefont {A.}~\bibnamefont
  {Srivastava}}\ and\ \bibinfo {author} {\bibfnamefont {S.}~\bibnamefont
  {Mukerjee}},\ }\href {https://doi.org/10.48550/ARXIV.2511.17088} {\bibinfo
  {title} {Electron {Hydrodynamics}: {Viscosity} {Tensor} and effects of a
  {Magnetic} field}} (\bibinfo {year} {2025}),\ \bibinfo {note}
  {arXiv:2511.17088 [cond-mat.str-el]}\BibitemShut {NoStop}%
\bibitem [{\citenamefont {Banerjee}\ \emph {et~al.}(2017)\citenamefont
  {Banerjee}, \citenamefont {Souslov}, \citenamefont {Abanov},\ and\
  \citenamefont {Vitelli}}]{banerjee_odd_2017}%
  \BibitemOpen
  \bibfield  {author} {\bibinfo {author} {\bibfnamefont {D.}~\bibnamefont
  {Banerjee}}, \bibinfo {author} {\bibfnamefont {A.}~\bibnamefont {Souslov}},
  \bibinfo {author} {\bibfnamefont {A.~G.}\ \bibnamefont {Abanov}},\ and\
  \bibinfo {author} {\bibfnamefont {V.}~\bibnamefont {Vitelli}},\ }\bibfield
  {title} {\bibinfo {title} {Odd viscosity in chiral active fluids},\ }\href
  {https://doi.org/10.1038/s41467-017-01378-7} {\bibfield  {journal} {\bibinfo
  {journal} {Nat Commun}\ }\textbf {\bibinfo {volume} {8}},\ \bibinfo {pages}
  {1573} (\bibinfo {year} {2017})}\BibitemShut {NoStop}%
\bibitem [{\citenamefont {Soni}\ \emph {et~al.}(2019)\citenamefont {Soni},
  \citenamefont {Bililign}, \citenamefont {Magkiriadou}, \citenamefont
  {Sacanna}, \citenamefont {Bartolo}, \citenamefont {Shelley},\ and\
  \citenamefont {Irvine}}]{soni_odd_2019}%
  \BibitemOpen
  \bibfield  {author} {\bibinfo {author} {\bibfnamefont {V.}~\bibnamefont
  {Soni}}, \bibinfo {author} {\bibfnamefont {E.~S.}\ \bibnamefont {Bililign}},
  \bibinfo {author} {\bibfnamefont {S.}~\bibnamefont {Magkiriadou}}, \bibinfo
  {author} {\bibfnamefont {S.}~\bibnamefont {Sacanna}}, \bibinfo {author}
  {\bibfnamefont {D.}~\bibnamefont {Bartolo}}, \bibinfo {author} {\bibfnamefont
  {M.~J.}\ \bibnamefont {Shelley}},\ and\ \bibinfo {author} {\bibfnamefont
  {W.~T.~M.}\ \bibnamefont {Irvine}},\ }\bibfield  {title} {\bibinfo {title}
  {The odd free surface flows of a colloidal chiral fluid},\ }\href
  {https://doi.org/10.1038/s41567-019-0603-8} {\bibfield  {journal} {\bibinfo
  {journal} {Nat. Phys.}\ }\textbf {\bibinfo {volume} {15}},\ \bibinfo {pages}
  {1188} (\bibinfo {year} {2019})}\BibitemShut {NoStop}%
\bibitem [{\citenamefont {Markovich}\ and\ \citenamefont
  {Lubensky}(2021)}]{markovich_odd_2021}%
  \BibitemOpen
  \bibfield  {author} {\bibinfo {author} {\bibfnamefont {T.}~\bibnamefont
  {Markovich}}\ and\ \bibinfo {author} {\bibfnamefont {T.~C.}\ \bibnamefont
  {Lubensky}},\ }\bibfield  {title} {\bibinfo {title} {Odd {Viscosity} in
  {Active} {Matter}: {Microscopic} {Origin} and {3D} {Effects}},\ }\href
  {https://doi.org/10.1103/PhysRevLett.127.048001} {\bibfield  {journal}
  {\bibinfo  {journal} {Phys. Rev. Lett.}\ }\textbf {\bibinfo {volume} {127}},\
  \bibinfo {pages} {048001} (\bibinfo {year} {2021})}\BibitemShut {NoStop}%
\bibitem [{\citenamefont {Berdyugin}\ \emph {et~al.}(2019)\citenamefont
  {Berdyugin}, \citenamefont {Xu}, \citenamefont {Pellegrino}, \citenamefont
  {Krishna~Kumar}, \citenamefont {Principi}, \citenamefont {Torre},
  \citenamefont {Ben~Shalom}, \citenamefont {Taniguchi}, \citenamefont
  {Watanabe}, \citenamefont {Grigorieva}, \citenamefont {Polini}, \citenamefont
  {Geim},\ and\ \citenamefont {Bandurin}}]{berdyugin_measuring_2019}%
  \BibitemOpen
  \bibfield  {author} {\bibinfo {author} {\bibfnamefont {A.~I.}\ \bibnamefont
  {Berdyugin}}, \bibinfo {author} {\bibfnamefont {S.~G.}\ \bibnamefont {Xu}},
  \bibinfo {author} {\bibfnamefont {F.~M.~D.}\ \bibnamefont {Pellegrino}},
  \bibinfo {author} {\bibfnamefont {R.}~\bibnamefont {Krishna~Kumar}}, \bibinfo
  {author} {\bibfnamefont {A.}~\bibnamefont {Principi}}, \bibinfo {author}
  {\bibfnamefont {I.}~\bibnamefont {Torre}}, \bibinfo {author} {\bibfnamefont
  {M.}~\bibnamefont {Ben~Shalom}}, \bibinfo {author} {\bibfnamefont
  {T.}~\bibnamefont {Taniguchi}}, \bibinfo {author} {\bibfnamefont
  {K.}~\bibnamefont {Watanabe}}, \bibinfo {author} {\bibfnamefont {I.~V.}\
  \bibnamefont {Grigorieva}}, \bibinfo {author} {\bibfnamefont
  {M.}~\bibnamefont {Polini}}, \bibinfo {author} {\bibfnamefont {A.~K.}\
  \bibnamefont {Geim}},\ and\ \bibinfo {author} {\bibfnamefont {D.~A.}\
  \bibnamefont {Bandurin}},\ }\bibfield  {title} {\bibinfo {title} {Measuring
  {Hall} viscosity of graphene's electron fluid},\ }\href
  {https://doi.org/10.1126/science.aau0685} {\bibfield  {journal} {\bibinfo
  {journal} {Science}\ }\textbf {\bibinfo {volume} {364}},\ \bibinfo {pages}
  {162} (\bibinfo {year} {2019})}\BibitemShut {NoStop}%
\bibitem [{\citenamefont {Li}\ \emph {et~al.}(2026)\citenamefont {Li},
  \citenamefont {Qin},\ and\ \citenamefont {Zhou}}]{li_electronic_2026}%
  \BibitemOpen
  \bibfield  {author} {\bibinfo {author} {\bibfnamefont {D.}~\bibnamefont
  {Li}}, \bibinfo {author} {\bibfnamefont {T.}~\bibnamefont {Qin}},\ and\
  \bibinfo {author} {\bibfnamefont {J.}~\bibnamefont {Zhou}},\ }\href
  {https://doi.org/10.48550/arXiv.2606.01003} {\bibinfo {title} {Electronic
  {Hall} viscosity: hidden indicator for antiferromagnets}} (\bibinfo {year}
  {2026}),\ \bibinfo {note} {arXiv:2606.01003 [cond-mat.mes-hall]}\BibitemShut
  {NoStop}%
\bibitem [{\citenamefont {Drazin}\ and\ \citenamefont
  {Reid}(2004)}]{drazin_hydrodynamic_2004}%
  \BibitemOpen
  \bibfield  {author} {\bibinfo {author} {\bibfnamefont {P.~G.}\ \bibnamefont
  {Drazin}}\ and\ \bibinfo {author} {\bibfnamefont {W.~H.}\ \bibnamefont
  {Reid}},\ }\href@noop {} {\emph {\bibinfo {title} {Hydrodynamic
  stability}}},\ \bibinfo {edition} {2nd}\ ed.,\ Cambridge monographs on
  mechanics and applied mathematics\ (\bibinfo  {publisher} {Cambridge
  University Press},\ \bibinfo {address} {Cambridge, UK ; New York},\ \bibinfo
  {year} {2004})\BibitemShut {NoStop}%
\bibitem [{\citenamefont {Kirkinis}\ and\ \citenamefont
  {Andreev}(2019)}]{kirkinis_odd-viscosity-induced_2019}%
  \BibitemOpen
  \bibfield  {author} {\bibinfo {author} {\bibfnamefont {E.}~\bibnamefont
  {Kirkinis}}\ and\ \bibinfo {author} {\bibfnamefont {A.~V.}\ \bibnamefont
  {Andreev}},\ }\bibfield  {title} {\bibinfo {title} {Odd-viscosity-induced
  stabilization of viscous thin liquid films},\ }\href
  {https://doi.org/10.1017/jfm.2019.644} {\bibfield  {journal} {\bibinfo
  {journal} {J. Fluid Mech.}\ }\textbf {\bibinfo {volume} {878}},\ \bibinfo
  {pages} {169} (\bibinfo {year} {2019})}\BibitemShut {NoStop}%
\bibitem [{\citenamefont {Mukhopadhyay}\ and\ \citenamefont
  {Mukhopadhyay}(2021)}]{mukhopadhyay_thermocapillary_2021}%
  \BibitemOpen
  \bibfield  {author} {\bibinfo {author} {\bibfnamefont {S.}~\bibnamefont
  {Mukhopadhyay}}\ and\ \bibinfo {author} {\bibfnamefont {A.}~\bibnamefont
  {Mukhopadhyay}},\ }\bibfield  {title} {\bibinfo {title} {Thermocapillary
  instability and wave formation on a viscous film flowing down an inclined
  plane with linear temperature variation: {Effect} of odd viscosity},\ }\href
  {https://doi.org/10.1063/5.0040260} {\bibfield  {journal} {\bibinfo
  {journal} {Physics of Fluids}\ }\textbf {\bibinfo {volume} {33}},\ \bibinfo
  {pages} {034110} (\bibinfo {year} {2021})}\BibitemShut {NoStop}%
\bibitem [{\citenamefont {Zhao}\ and\ \citenamefont
  {Jian}(2021)}]{zhao_effect_2021}%
  \BibitemOpen
  \bibfield  {author} {\bibinfo {author} {\bibfnamefont {J.}~\bibnamefont
  {Zhao}}\ and\ \bibinfo {author} {\bibfnamefont {Y.}~\bibnamefont {Jian}},\
  }\bibfield  {title} {\bibinfo {title} {Effect of odd viscosity on the
  stability of a falling thin film in presence of electromagnetic field},\
  }\href {https://doi.org/10.1088/1873-7005/abde23} {\bibfield  {journal}
  {\bibinfo  {journal} {Fluid Dyn. Res.}\ }\textbf {\bibinfo {volume} {53}},\
  \bibinfo {pages} {015510} (\bibinfo {year} {2021})}\BibitemShut {NoStop}%
\bibitem [{\citenamefont {Samanta}(2022)}]{samanta_role_2022}%
  \BibitemOpen
  \bibfield  {author} {\bibinfo {author} {\bibfnamefont {A.}~\bibnamefont
  {Samanta}},\ }\bibfield  {title} {\bibinfo {title} {Role of odd viscosity in
  falling viscous fluid},\ }\href {https://doi.org/10.1017/jfm.2022.155}
  {\bibfield  {journal} {\bibinfo  {journal} {J. Fluid Mech.}\ }\textbf
  {\bibinfo {volume} {938}},\ \bibinfo {pages} {A9} (\bibinfo {year}
  {2022})}\BibitemShut {NoStop}%
\bibitem [{\citenamefont {Hossain}\ \emph {et~al.}(2025)\citenamefont
  {Hossain}, \citenamefont {Saha}, \citenamefont {Ghosh},\ and\ \citenamefont
  {Behera}}]{hossain_stability_2025}%
  \BibitemOpen
  \bibfield  {author} {\bibinfo {author} {\bibfnamefont {M.~M.}\ \bibnamefont
  {Hossain}}, \bibinfo {author} {\bibfnamefont {M.}~\bibnamefont {Saha}},
  \bibinfo {author} {\bibfnamefont {S.}~\bibnamefont {Ghosh}},\ and\ \bibinfo
  {author} {\bibfnamefont {H.}~\bibnamefont {Behera}},\ }\bibfield  {title}
  {\bibinfo {title} {Stability dynamics of an odd-viscosity induced fluid flow
  over a vibrating inclined surface},\ }\href
  {https://doi.org/10.1063/5.0291294} {\bibfield  {journal} {\bibinfo
  {journal} {Physics of Fluids}\ }\textbf {\bibinfo {volume} {37}},\ \bibinfo
  {pages} {104108} (\bibinfo {year} {2025})}\BibitemShut {NoStop}%
\bibitem [{\citenamefont {Tong}\ \emph {et~al.}(2026)\citenamefont {Tong},
  \citenamefont {Xie},\ and\ \citenamefont {Jian}}]{tong_influence_2026}%
  \BibitemOpen
  \bibfield  {author} {\bibinfo {author} {\bibfnamefont {J.}~\bibnamefont
  {Tong}}, \bibinfo {author} {\bibfnamefont {Z.}~\bibnamefont {Xie}},\ and\
  \bibinfo {author} {\bibfnamefont {Y.}~\bibnamefont {Jian}},\ }\bibfield
  {title} {\bibinfo {title} {Influence of the odd viscosity on the instability
  of a horizontal oscillating thin liquid film},\ }\href
  {https://doi.org/10.1063/5.0314895} {\bibfield  {journal} {\bibinfo
  {journal} {Physics of Fluids}\ }\textbf {\bibinfo {volume} {38}},\ \bibinfo
  {pages} {034101} (\bibinfo {year} {2026})}\BibitemShut {NoStop}%
\bibitem [{\citenamefont {Steinhauer}\ and\ \citenamefont
  {Ishida}(1990)}]{steinhauer_gyroviscous_1990}%
  \BibitemOpen
  \bibfield  {author} {\bibinfo {author} {\bibfnamefont {L.~C.}\ \bibnamefont
  {Steinhauer}}\ and\ \bibinfo {author} {\bibfnamefont {A.}~\bibnamefont
  {Ishida}},\ }\bibfield  {title} {\bibinfo {title} {Gyroviscous stability
  theory with application to the internal tilt mode of a field-reversed
  configuration},\ }\href {https://doi.org/10.1063/1.859507} {\bibfield
  {journal} {\bibinfo  {journal} {Physics of Fluids B: Plasma Physics}\
  }\textbf {\bibinfo {volume} {2}},\ \bibinfo {pages} {2422} (\bibinfo {year}
  {1990})}\BibitemShut {NoStop}%
\bibitem [{\citenamefont {Du}\ and\ \citenamefont
  {Podgornik}(2023)}]{du_stability_2023}%
  \BibitemOpen
  \bibfield  {author} {\bibinfo {author} {\bibfnamefont {G.}~\bibnamefont
  {Du}}\ and\ \bibinfo {author} {\bibfnamefont {R.}~\bibnamefont {Podgornik}},\
  }\href {https://doi.org/10.48550/arXiv.2309.09594} {\bibinfo {title}
  {Stability of {Taylor}-{Couette} {Flow} with {Odd} {Viscosity}}} (\bibinfo
  {year} {2023}),\ \bibinfo {note} {arXiv:2309.09594 [physics]}\BibitemShut
  {NoStop}%
\bibitem [{\citenamefont {de~Wit}\ \emph {et~al.}(2024)\citenamefont {de~Wit},
  \citenamefont {Fruchart}, \citenamefont {Khain}, \citenamefont {Toschi},\
  and\ \citenamefont {Vitelli}}]{de_wit_pattern_2024}%
  \BibitemOpen
  \bibfield  {author} {\bibinfo {author} {\bibfnamefont {X.~M.}\ \bibnamefont
  {de~Wit}}, \bibinfo {author} {\bibfnamefont {M.}~\bibnamefont {Fruchart}},
  \bibinfo {author} {\bibfnamefont {T.}~\bibnamefont {Khain}}, \bibinfo
  {author} {\bibfnamefont {F.}~\bibnamefont {Toschi}},\ and\ \bibinfo {author}
  {\bibfnamefont {V.}~\bibnamefont {Vitelli}},\ }\bibfield  {title} {\bibinfo
  {title} {Pattern formation by turbulent cascades},\ }\href
  {https://doi.org/10.1038/s41586-024-07074-z} {\bibfield  {journal} {\bibinfo
  {journal} {Nature}\ }\textbf {\bibinfo {volume} {627}},\ \bibinfo {pages}
  {515} (\bibinfo {year} {2024})}\BibitemShut {NoStop}%
\bibitem [{\citenamefont {Chen}\ \emph {et~al.}(2024)\citenamefont {Chen},
  \citenamefont {de~Wit}, \citenamefont {Fruchart}, \citenamefont {Toschi},\
  and\ \citenamefont {Vitelli}}]{chen_odd_2024}%
  \BibitemOpen
  \bibfield  {author} {\bibinfo {author} {\bibfnamefont {S.}~\bibnamefont
  {Chen}}, \bibinfo {author} {\bibfnamefont {X.~M.}\ \bibnamefont {de~Wit}},
  \bibinfo {author} {\bibfnamefont {M.}~\bibnamefont {Fruchart}}, \bibinfo
  {author} {\bibfnamefont {F.}~\bibnamefont {Toschi}},\ and\ \bibinfo {author}
  {\bibfnamefont {V.}~\bibnamefont {Vitelli}},\ }\bibfield  {title} {\bibinfo
  {title} {Odd {Viscosity} {Suppresses} {Intermittency} in {Direct} {Turbulent}
  {Cascades}},\ }\href {https://doi.org/10.1103/PhysRevLett.133.144002}
  {\bibfield  {journal} {\bibinfo  {journal} {Phys. Rev. Lett.}\ }\textbf
  {\bibinfo {volume} {133}},\ \bibinfo {pages} {144002} (\bibinfo {year}
  {2024})}\BibitemShut {NoStop}%
\bibitem [{\citenamefont {de~Wit}\ \emph {et~al.}(2026)\citenamefont {de~Wit},
  \citenamefont {Touzo}, \citenamefont {Galtier}, \citenamefont {Fruchart},
  \citenamefont {Toschi},\ and\ \citenamefont {Vitelli}}]{de_wit_wave_2026}%
  \BibitemOpen
  \bibfield  {author} {\bibinfo {author} {\bibfnamefont {X.~M.}\ \bibnamefont
  {de~Wit}}, \bibinfo {author} {\bibfnamefont {L.}~\bibnamefont {Touzo}},
  \bibinfo {author} {\bibfnamefont {S.}~\bibnamefont {Galtier}}, \bibinfo
  {author} {\bibfnamefont {M.}~\bibnamefont {Fruchart}}, \bibinfo {author}
  {\bibfnamefont {F.}~\bibnamefont {Toschi}},\ and\ \bibinfo {author}
  {\bibfnamefont {V.}~\bibnamefont {Vitelli}},\ }\href
  {https://doi.org/10.48550/ARXIV.2606.14583} {\bibinfo {title} {Wave
  turbulence theory of odd fluids and solids: kinetic equations and solutions}}
  (\bibinfo {year} {2026}),\ \bibinfo {note} {arXiv:2606.14583
  [physics.flu-dyn]}\BibitemShut {NoStop}%
\bibitem [{\citenamefont {Schmid}\ and\ \citenamefont
  {Henningson}(2001)}]{schmid_stability_2001}%
  \BibitemOpen
  \bibfield  {author} {\bibinfo {author} {\bibfnamefont {P.~J.}\ \bibnamefont
  {Schmid}}\ and\ \bibinfo {author} {\bibfnamefont {D.~S.}\ \bibnamefont
  {Henningson}},\ }\href {https://doi.org/10.1007/978-1-4613-0185-1} {\emph
  {\bibinfo {title} {Stability and {Transition} in {Shear} {Flows}}}},\ edited
  by\ \bibinfo {editor} {\bibfnamefont {J.~E.}\ \bibnamefont {Marsden}}\ and\
  \bibinfo {editor} {\bibfnamefont {L.}~\bibnamefont {Sirovich}},\ \bibinfo
  {series} {Applied {Mathematical} {Sciences}}, Vol.\ \bibinfo {volume} {142}\
  (\bibinfo  {publisher} {Springer New York},\ \bibinfo {address} {New York,
  NY},\ \bibinfo {year} {2001})\BibitemShut {NoStop}%
\bibitem [{\citenamefont {Reynolds}\ \emph {et~al.}(2023)\citenamefont
  {Reynolds}, \citenamefont {Monteiro},\ and\ \citenamefont
  {Ganeshan}}]{reynolds_three_2023}%
  \BibitemOpen
  \bibfield  {author} {\bibinfo {author} {\bibfnamefont {D.}~\bibnamefont
  {Reynolds}}, \bibinfo {author} {\bibfnamefont {G.~M.}\ \bibnamefont
  {Monteiro}},\ and\ \bibinfo {author} {\bibfnamefont {S.}~\bibnamefont
  {Ganeshan}},\ }\href {https://doi.org/10.48550/ARXIV.2301.07096} {\bibinfo
  {title} {Three {Dimensional} {Odd} {Viscosity} in {Ferrofluids} with
  {Vorticity}-{Magnetization} {Coupling}}} (\bibinfo {year} {2023}),\ \bibinfo
  {note} {arXiv:2301.07096 [physics.flu-dyn]}\BibitemShut {NoStop}%
\bibitem [{\citenamefont {Ganeshan}\ and\ \citenamefont
  {Abanov}(2017)}]{ganeshan_odd_2017}%
  \BibitemOpen
  \bibfield  {author} {\bibinfo {author} {\bibfnamefont {S.}~\bibnamefont
  {Ganeshan}}\ and\ \bibinfo {author} {\bibfnamefont {A.~G.}\ \bibnamefont
  {Abanov}},\ }\bibfield  {title} {\bibinfo {title} {Odd viscosity in
  two-dimensional incompressible fluids},\ }\href
  {https://doi.org/10.1103/PhysRevFluids.2.094101} {\bibfield  {journal}
  {\bibinfo  {journal} {Phys. Rev. Fluids}\ }\textbf {\bibinfo {volume} {2}},\
  \bibinfo {pages} {094101} (\bibinfo {year} {2017})}\BibitemShut {NoStop}%
\bibitem [{\citenamefont {Chapman}(2002)}]{chapman_subcritical_2002}%
  \BibitemOpen
  \bibfield  {author} {\bibinfo {author} {\bibfnamefont {S.~J.}\ \bibnamefont
  {Chapman}},\ }\bibfield  {title} {\bibinfo {title} {Subcritical transition in
  channel flows},\ }\href {https://doi.org/10.1017/S0022112001006255}
  {\bibfield  {journal} {\bibinfo  {journal} {J. Fluid Mech.}\ }\textbf
  {\bibinfo {volume} {451}},\ \bibinfo {pages} {35} (\bibinfo {year}
  {2002})}\BibitemShut {NoStop}%
\bibitem [{\citenamefont {Berry}(2004)}]{berry_physics_2004}%
  \BibitemOpen
  \bibfield  {author} {\bibinfo {author} {\bibfnamefont {M.~V.}\ \bibnamefont
  {Berry}},\ }\bibfield  {title} {\bibinfo {title} {Physics of {Nonhermitian}
  {Degeneracies}},\ }\href {https://doi.org/10.1023/B:CJOP.0000044002.05657.04}
  {\bibfield  {journal} {\bibinfo  {journal} {Czechoslovak Journal of Physics}\
  }\textbf {\bibinfo {volume} {54}},\ \bibinfo {pages} {1039} (\bibinfo {year}
  {2004})}\BibitemShut {NoStop}%
\bibitem [{\citenamefont {Heiss}(2004)}]{heiss_exceptional_2004}%
  \BibitemOpen
  \bibfield  {author} {\bibinfo {author} {\bibfnamefont {W.~D.}\ \bibnamefont
  {Heiss}},\ }\bibfield  {title} {\bibinfo {title} {Exceptional points of
  non-{Hermitian} operators},\ }\href
  {https://doi.org/10.1088/0305-4470/37/6/034} {\bibfield  {journal} {\bibinfo
  {journal} {J. Phys. A: Math. Gen.}\ }\textbf {\bibinfo {volume} {37}},\
  \bibinfo {pages} {2455} (\bibinfo {year} {2004})}\BibitemShut {NoStop}%
\bibitem [{\citenamefont {Kern}\ \emph {et~al.}(2022)\citenamefont {Kern},
  \citenamefont {Hanifi},\ and\ \citenamefont
  {Henningson}}]{kern_subharmonic_2022}%
  \BibitemOpen
  \bibfield  {author} {\bibinfo {author} {\bibfnamefont {J.~S.}\ \bibnamefont
  {Kern}}, \bibinfo {author} {\bibfnamefont {A.}~\bibnamefont {Hanifi}},\ and\
  \bibinfo {author} {\bibfnamefont {D.~S.}\ \bibnamefont {Henningson}},\
  }\bibfield  {title} {\bibinfo {title} {Subharmonic eigenvalue orbits in the
  spectrum of pulsating {Poiseuille} flow},\ }\href
  {https://doi.org/10.1017/jfm.2022.515} {\bibfield  {journal} {\bibinfo
  {journal} {J. Fluid Mech.}\ }\textbf {\bibinfo {volume} {945}},\ \bibinfo
  {pages} {A11} (\bibinfo {year} {2022})}\BibitemShut {NoStop}%
\bibitem [{\citenamefont {Kirkinis}\ and\ \citenamefont
  {Levchenko}(2026)}]{kirkinis_mathscrpt-symmetric_2026}%
  \BibitemOpen
  \bibfield  {author} {\bibinfo {author} {\bibfnamefont {E.}~\bibnamefont
  {Kirkinis}}\ and\ \bibinfo {author} {\bibfnamefont {A.}~\bibnamefont
  {Levchenko}},\ }\href {https://doi.org/10.48550/arXiv.2608.22958} {\bibinfo
  {title} {PT-symmetric hydrodynamics of odd viscous liquids and
  their oscillator counterparts}} (\bibinfo {year} {2026}),\ \bibinfo {note}
  {arXiv:2608.22958 [physics.flu-dyn]}\BibitemShut {NoStop}%
\bibitem [{\citenamefont {Lezius}\ and\ \citenamefont
  {Johnston}(1976)}]{lezius_roll-cell_1976}%
  \BibitemOpen
  \bibfield  {author} {\bibinfo {author} {\bibfnamefont {D.~K.}\ \bibnamefont
  {Lezius}}\ and\ \bibinfo {author} {\bibfnamefont {J.~P.}\ \bibnamefont
  {Johnston}},\ }\bibfield  {title} {\bibinfo {title} {Roll-cell instabilities
  in rotating laminar and trubulent channel flows},\ }\href
  {https://doi.org/10.1017/S0022112076001171} {\bibfield  {journal} {\bibinfo
  {journal} {J. Fluid Mech.}\ }\textbf {\bibinfo {volume} {77}},\ \bibinfo
  {pages} {153} (\bibinfo {year} {1976})}\BibitemShut {NoStop}%
\bibitem [{\citenamefont {Alfredsson}\ and\ \citenamefont
  {Persson}(1989)}]{alfredsson_instabilities_1989}%
  \BibitemOpen
  \bibfield  {author} {\bibinfo {author} {\bibfnamefont {P.~H.}\ \bibnamefont
  {Alfredsson}}\ and\ \bibinfo {author} {\bibfnamefont {H.}~\bibnamefont
  {Persson}},\ }\bibfield  {title} {\bibinfo {title} {Instabilities in channel
  flow with system rotation},\ }\href
  {https://doi.org/10.1017/S002211208900128X} {\bibfield  {journal} {\bibinfo
  {journal} {J. Fluid Mech.}\ }\textbf {\bibinfo {volume} {202}},\ \bibinfo
  {pages} {543} (\bibinfo {year} {1989})}\BibitemShut {NoStop}%
\bibitem [{\citenamefont {Wall}\ and\ \citenamefont
  {Nagata}(2006)}]{wall_nonlinear_2006}%
  \BibitemOpen
  \bibfield  {author} {\bibinfo {author} {\bibfnamefont {D.~P.}\ \bibnamefont
  {Wall}}\ and\ \bibinfo {author} {\bibfnamefont {M.}~\bibnamefont {Nagata}},\
  }\bibfield  {title} {\bibinfo {title} {Nonlinear secondary flow through a
  rotating channel},\ }\href {https://doi.org/10.1017/S0022112006001157}
  {\bibfield  {journal} {\bibinfo  {journal} {J. Fluid Mech.}\ }\textbf
  {\bibinfo {volume} {564}},\ \bibinfo {pages} {25} (\bibinfo {year}
  {2006})}\BibitemShut {NoStop}%
\bibitem [{\citenamefont {Brethouwer}(2025)}]{brethouwer_stability_2025}%
  \BibitemOpen
  \bibfield  {author} {\bibinfo {author} {\bibfnamefont {G.}~\bibnamefont
  {Brethouwer}},\ }\bibfield  {title} {\bibinfo {title} {Stability of plane
  {Couette} and {Poiseuille} flows rotating about the streamwise axis},\ }\href
  {https://doi.org/10.1017/jfm.2025.10723} {\bibfield  {journal} {\bibinfo
  {journal} {J. Fluid Mech.}\ }\textbf {\bibinfo {volume} {1021}},\ \bibinfo
  {pages} {A14} (\bibinfo {year} {2025})}\BibitemShut {NoStop}%
\bibitem [{\citenamefont {Kirkinis}\ and\ \citenamefont {Olvera de~la
  Cruz}(2023)}]{kirkinis_taylor_2023}%
  \BibitemOpen
  \bibfield  {author} {\bibinfo {author} {\bibfnamefont {E.}~\bibnamefont
  {Kirkinis}}\ and\ \bibinfo {author} {\bibfnamefont {M.}~\bibnamefont {Olvera
  de~la Cruz}},\ }\bibfield  {title} {\bibinfo {title} {Taylor columns and
  inertial-like waves in a three-dimensional odd viscous liquid},\ }\href
  {https://doi.org/10.1017/jfm.2023.769} {\bibfield  {journal} {\bibinfo
  {journal} {J. Fluid Mech.}\ }\textbf {\bibinfo {volume} {973}},\ \bibinfo
  {pages} {A30} (\bibinfo {year} {2023})}\BibitemShut {NoStop}%
\bibitem [{\citenamefont
  {Chandrasekhar}(1961)}]{chandrasekhar_hydrodynamic_1961}%
  \BibitemOpen
  \bibfield  {author} {\bibinfo {author} {\bibfnamefont {S.}~\bibnamefont
  {Chandrasekhar}},\ }\href@noop {} {\emph {\bibinfo {title} {Hydrodynamic and
  {Hydromagnetic} {Stability}}}},\ International series of monographs on
  physics\ (\bibinfo  {publisher} {Clarendon Pr.},\ \bibinfo {address}
  {Oxford},\ \bibinfo {year} {1961})\BibitemShut {NoStop}%
\bibitem [{\citenamefont {Herron}(2001)}]{herron_principle_2001}%
  \BibitemOpen
  \bibfield  {author} {\bibinfo {author} {\bibfnamefont {I.~H.}\ \bibnamefont
  {Herron}},\ }\bibfield  {title} {\bibinfo {title} {On the {Principle} of
  {Exchange} of {Stabilities} in {Rayleigh}--{B{\'e}nard} {Convection}},\
  }\href {https://doi.org/10.1137/S0036139900370388} {\bibfield  {journal}
  {\bibinfo  {journal} {SIAM J. Appl. Math.}\ }\textbf {\bibinfo {volume}
  {61}},\ \bibinfo {pages} {1362} (\bibinfo {year} {2001})}\BibitemShut
  {NoStop}%
\bibitem [{\citenamefont {Herron}\ and\ \citenamefont
  {Ali}(2003)}]{herron_principle_2003}%
  \BibitemOpen
  \bibfield  {author} {\bibinfo {author} {\bibfnamefont {I.~H.}\ \bibnamefont
  {Herron}}\ and\ \bibinfo {author} {\bibfnamefont {H.~N.}\ \bibnamefont
  {Ali}},\ }\bibfield  {title} {\bibinfo {title} {The principle of exchange of
  stabilities for {Couette} flow},\ }\href
  {https://doi.org/10.1090/qam/1976370} {\bibfield  {journal} {\bibinfo
  {journal} {Quart. Appl. Math.}\ }\textbf {\bibinfo {volume} {61}},\ \bibinfo
  {pages} {279} (\bibinfo {year} {2003})}\BibitemShut {NoStop}%
\bibitem [{\citenamefont {Jose}\ and\ \citenamefont
  {Govindarajan}(2020)}]{jose_non_normal_2020}%
  \BibitemOpen
  \bibfield  {author} {\bibinfo {author} {\bibfnamefont {S.}~\bibnamefont
  {Jose}}\ and\ \bibinfo {author} {\bibfnamefont {R.}~\bibnamefont
  {Govindarajan}},\ }\bibfield  {title} {\bibinfo {title} {Non-normal origin of
  modal instabilities in rotating plane shear flows},\ }\href
  {https://doi.org/10.1098/rspa.2019.0550} {\bibfield  {journal} {\bibinfo
  {journal} {Proc. R. Soc. A.}\ }\textbf {\bibinfo {volume} {476}},\ \bibinfo
  {pages} {20190550} (\bibinfo {year} {2020})}\BibitemShut {NoStop}%
\bibitem [{Note1()}]{Note1}%
  \BibitemOpen
  \bibinfo {note} {We note Ref. \cite {jose_non_normal_2020} for a related
  discussion of the connection between rotating-flow instabilities and the
  pseudospectrum of the Navier--Stokes operator.}\BibitemShut {Stop}%
\bibitem [{\citenamefont {Reddy}\ \emph {et~al.}(1993)\citenamefont {Reddy},
  \citenamefont {Schmid},\ and\ \citenamefont
  {Henningson}}]{reddy_pseudospectra_1993}%
  \BibitemOpen
  \bibfield  {author} {\bibinfo {author} {\bibfnamefont {S.~C.}\ \bibnamefont
  {Reddy}}, \bibinfo {author} {\bibfnamefont {P.~J.}\ \bibnamefont {Schmid}},\
  and\ \bibinfo {author} {\bibfnamefont {D.~S.}\ \bibnamefont {Henningson}},\
  }\bibfield  {title} {\bibinfo {title} {Pseudospectra of the
  {Orr}--{Sommerfeld} {Operator}},\ }\href {https://doi.org/10.1137/0153002}
  {\bibfield  {journal} {\bibinfo  {journal} {SIAM J. Appl. Math.}\ }\textbf
  {\bibinfo {volume} {53}},\ \bibinfo {pages} {15} (\bibinfo {year}
  {1993})}\BibitemShut {NoStop}%
\bibitem [{\citenamefont {Schmid}(2007)}]{schmid_nonmodal_2007}%
  \BibitemOpen
  \bibfield  {author} {\bibinfo {author} {\bibfnamefont {P.~J.}\ \bibnamefont
  {Schmid}},\ }\bibfield  {title} {\bibinfo {title} {Nonmodal {Stability}
  {Theory}},\ }\href {https://doi.org/10.1146/annurev.fluid.38.050304.092139}
  {\bibfield  {journal} {\bibinfo  {journal} {Annu. Rev. Fluid Mech.}\ }\textbf
  {\bibinfo {volume} {39}},\ \bibinfo {pages} {129} (\bibinfo {year}
  {2007})}\BibitemShut {NoStop}%
\bibitem [{\citenamefont {Butler}\ and\ \citenamefont
  {Farrell}(1992)}]{butler_three-dimensional_1992}%
  \BibitemOpen
  \bibfield  {author} {\bibinfo {author} {\bibfnamefont {K.~M.}\ \bibnamefont
  {Butler}}\ and\ \bibinfo {author} {\bibfnamefont {B.~F.}\ \bibnamefont
  {Farrell}},\ }\bibfield  {title} {\bibinfo {title} {Three-dimensional optimal
  perturbations in viscous shear flow},\ }\href
  {https://doi.org/10.1063/1.858386} {\bibfield  {journal} {\bibinfo  {journal}
  {Physics of Fluids A: Fluid Dynamics}\ }\textbf {\bibinfo {volume} {4}},\
  \bibinfo {pages} {1637} (\bibinfo {year} {1992})}\BibitemShut {NoStop}%
\bibitem [{\citenamefont {Carlson}\ \emph {et~al.}(1982)\citenamefont
  {Carlson}, \citenamefont {Widnall},\ and\ \citenamefont
  {Peeters}}]{carlson_flow_visualization_1982}%
  \BibitemOpen
  \bibfield  {author} {\bibinfo {author} {\bibfnamefont {D.~R.}\ \bibnamefont
  {Carlson}}, \bibinfo {author} {\bibfnamefont {S.~E.}\ \bibnamefont
  {Widnall}},\ and\ \bibinfo {author} {\bibfnamefont {M.~F.}\ \bibnamefont
  {Peeters}},\ }\bibfield  {title} {\bibinfo {title} {A flow-visualization
  study of transition in plane {Poiseuille} flow},\ }\href
  {https://doi.org/10.1017/S0022112082002006} {\bibfield  {journal} {\bibinfo
  {journal} {J. Fluid Mech.}\ }\textbf {\bibinfo {volume} {121}},\ \bibinfo
  {pages} {487} (\bibinfo {year} {1982})}\BibitemShut {NoStop}%
\bibitem [{\citenamefont {Lucas}\ and\ \citenamefont
  {Fong}(2018)}]{lucas_hydrodynamics_2018}%
  \BibitemOpen
  \bibfield  {author} {\bibinfo {author} {\bibfnamefont {A.}~\bibnamefont
  {Lucas}}\ and\ \bibinfo {author} {\bibfnamefont {K.~C.}\ \bibnamefont
  {Fong}},\ }\bibfield  {title} {\bibinfo {title} {Hydrodynamics of electrons
  in graphene},\ }\href {https://doi.org/10.1088/1361-648X/aaa274} {\bibfield
  {journal} {\bibinfo  {journal} {J. Phys.: Condens. Matter}\ }\textbf
  {\bibinfo {volume} {30}},\ \bibinfo {pages} {053001} (\bibinfo {year}
  {2018})}\BibitemShut {NoStop}%
\bibitem [{\citenamefont {Fritz}\ and\ \citenamefont
  {Scaffidi}(2024)}]{fritz_hydrodynamic_2024}%
  \BibitemOpen
  \bibfield  {author} {\bibinfo {author} {\bibfnamefont {L.}~\bibnamefont
  {Fritz}}\ and\ \bibinfo {author} {\bibfnamefont {T.}~\bibnamefont
  {Scaffidi}},\ }\bibfield  {title} {\bibinfo {title} {Hydrodynamic
  {Electronic} {Transport}},\ }\href
  {https://doi.org/10.1146/annurev-conmatphys-040521-042014} {\bibfield
  {journal} {\bibinfo  {journal} {Annu. Rev. Condens. Matter Phys.}\ }\textbf
  {\bibinfo {volume} {15}},\ \bibinfo {pages} {17} (\bibinfo {year}
  {2024})}\BibitemShut {NoStop}%
\bibitem [{\citenamefont {Hui}\ and\ \citenamefont
  {Skinner}(2025)}]{hui_hydrodynamics_2025}%
  \BibitemOpen
  \bibfield  {author} {\bibinfo {author} {\bibfnamefont {A.}~\bibnamefont
  {Hui}}\ and\ \bibinfo {author} {\bibfnamefont {B.}~\bibnamefont {Skinner}},\
  }\bibfield  {title} {\bibinfo {title} {Hydrodynamics of the electronic
  {Fermi} liquid: a pedagogical overview},\ }\href
  {https://doi.org/10.1088/1361-648X/adfbcd} {\bibfield  {journal} {\bibinfo
  {journal} {J. Phys.: Condens. Matter}\ }\textbf {\bibinfo {volume} {37}},\
  \bibinfo {pages} {363001} (\bibinfo {year} {2025})}\BibitemShut {NoStop}%
\bibitem [{\citenamefont {Zhang}\ \emph {et~al.}(2026)\citenamefont {Zhang},
  \citenamefont {Redekop}, \citenamefont {Stoyanov}, \citenamefont {Farrell},
  \citenamefont {Kim}, \citenamefont {Holleis}, \citenamefont {Gong},
  \citenamefont {Keough}, \citenamefont {Choi}, \citenamefont {Taniguchi},
  \citenamefont {Watanabe}, \citenamefont {Huber}, \citenamefont {{Bleszynski
  Jayich}}, \citenamefont {Lucas},\ and\ \citenamefont
  {Young}}]{zhang_imaging_2026}%
  \BibitemOpen
  \bibfield  {author} {\bibinfo {author} {\bibfnamefont {C.}~\bibnamefont
  {Zhang}}, \bibinfo {author} {\bibfnamefont {E.}~\bibnamefont {Redekop}},
  \bibinfo {author} {\bibfnamefont {H.}~\bibnamefont {Stoyanov}}, \bibinfo
  {author} {\bibfnamefont {J.~H.}\ \bibnamefont {Farrell}}, \bibinfo {author}
  {\bibfnamefont {S.}~\bibnamefont {Kim}}, \bibinfo {author} {\bibfnamefont
  {L.}~\bibnamefont {Holleis}}, \bibinfo {author} {\bibfnamefont
  {D.}~\bibnamefont {Gong}}, \bibinfo {author} {\bibfnamefont {A.}~\bibnamefont
  {Keough}}, \bibinfo {author} {\bibfnamefont {Y.}~\bibnamefont {Choi}},
  \bibinfo {author} {\bibfnamefont {T.}~\bibnamefont {Taniguchi}}, \bibinfo
  {author} {\bibfnamefont {K.}~\bibnamefont {Watanabe}}, \bibinfo {author}
  {\bibfnamefont {M.~E.}\ \bibnamefont {Huber}}, \bibinfo {author}
  {\bibfnamefont {A.~C.}\ \bibnamefont {{Bleszynski Jayich}}}, \bibinfo
  {author} {\bibfnamefont {A.}~\bibnamefont {Lucas}},\ and\ \bibinfo {author}
  {\bibfnamefont {A.~F.}\ \bibnamefont {Young}},\ }\href
  {https://doi.org/10.48550/ARXIV.2603.11175} {\bibinfo {title} {Imaging flat
  band electron hydrodynamics in biased bilayer graphene}} (\bibinfo {year}
  {2026}),\ \bibinfo {note} {arXiv:2603.11175 [cond-mat.mes-hall]}\BibitemShut
  {NoStop}%
\bibitem [{\citenamefont {Zhou}\ \emph {et~al.}(2021)\citenamefont {Zhou},
  \citenamefont {Xie}, \citenamefont {Ghazaryan}, \citenamefont {Holder},
  \citenamefont {Ehrets}, \citenamefont {Spanton}, \citenamefont {Taniguchi},
  \citenamefont {Watanabe}, \citenamefont {Berg}, \citenamefont {Serbyn},\ and\
  \citenamefont {Young}}]{zhou_half_2021}%
  \BibitemOpen
  \bibfield  {author} {\bibinfo {author} {\bibfnamefont {H.}~\bibnamefont
  {Zhou}}, \bibinfo {author} {\bibfnamefont {T.}~\bibnamefont {Xie}}, \bibinfo
  {author} {\bibfnamefont {A.}~\bibnamefont {Ghazaryan}}, \bibinfo {author}
  {\bibfnamefont {T.}~\bibnamefont {Holder}}, \bibinfo {author} {\bibfnamefont
  {J.~R.}\ \bibnamefont {Ehrets}}, \bibinfo {author} {\bibfnamefont {E.~M.}\
  \bibnamefont {Spanton}}, \bibinfo {author} {\bibfnamefont {T.}~\bibnamefont
  {Taniguchi}}, \bibinfo {author} {\bibfnamefont {K.}~\bibnamefont {Watanabe}},
  \bibinfo {author} {\bibfnamefont {E.}~\bibnamefont {Berg}}, \bibinfo {author}
  {\bibfnamefont {M.}~\bibnamefont {Serbyn}},\ and\ \bibinfo {author}
  {\bibfnamefont {A.~F.}\ \bibnamefont {Young}},\ }\bibfield  {title} {\bibinfo
  {title} {Half- and quarter-metals in rhombohedral trilayer graphene},\ }\href
  {https://doi.org/10.1038/s41586-021-03938-w} {\bibfield  {journal} {\bibinfo
  {journal} {Nature}\ }\textbf {\bibinfo {volume} {598}},\ \bibinfo {pages}
  {429} (\bibinfo {year} {2021})}\BibitemShut {NoStop}%
\bibitem [{\citenamefont {Xue}(2026)}]{xue_essay_2026}%
  \BibitemOpen
  \bibfield  {author} {\bibinfo {author} {\bibfnamefont {P.}~\bibnamefont
  {Xue}},\ }\bibfield  {title} {\bibinfo {title} {Essay: {Topological} {Phases}
  and {Exceptional} {Points} in {Non}-{Hermitian} {Systems}},\ }\href
  {https://doi.org/10.1103/ll76-j2l5} {\bibfield  {journal} {\bibinfo
  {journal} {Phys. Rev. Lett.}\ }\textbf {\bibinfo {volume} {136}},\ \bibinfo
  {pages} {170001} (\bibinfo {year} {2026})}\BibitemShut {NoStop}%
\bibitem [{\citenamefont {Khain}\ \emph {et~al.}(2022)\citenamefont {Khain},
  \citenamefont {Scheibner}, \citenamefont {Fruchart},\ and\ \citenamefont
  {Vitelli}}]{khain_stokes_2022}%
  \BibitemOpen
  \bibfield  {author} {\bibinfo {author} {\bibfnamefont {T.}~\bibnamefont
  {Khain}}, \bibinfo {author} {\bibfnamefont {C.}~\bibnamefont {Scheibner}},
  \bibinfo {author} {\bibfnamefont {M.}~\bibnamefont {Fruchart}},\ and\
  \bibinfo {author} {\bibfnamefont {V.}~\bibnamefont {Vitelli}},\ }\bibfield
  {title} {\bibinfo {title} {Stokes flows in three-dimensional fluids with odd
  and parity-violating viscosities},\ }\href
  {https://doi.org/10.1017/jfm.2021.1079} {\bibfield  {journal} {\bibinfo
  {journal} {J. Fluid Mech.}\ }\textbf {\bibinfo {volume} {934}},\ \bibinfo
  {pages} {A23} (\bibinfo {year} {2022})}\BibitemShut {NoStop}%
\bibitem [{\citenamefont {Kirkinis}\ and\ \citenamefont {Olvera de~la
  Cruz}(2024)}]{kirkinis_evanescent_2024}%
  \BibitemOpen
  \bibfield  {author} {\bibinfo {author} {\bibfnamefont {E.}~\bibnamefont
  {Kirkinis}}\ and\ \bibinfo {author} {\bibfnamefont {M.}~\bibnamefont {Olvera
  de~la Cruz}},\ }\bibfield  {title} {\bibinfo {title} {Evanescent and
  inertial-like waves in rigidly rotating odd viscous liquids},\ }\href
  {https://doi.org/10.1017/jfm.2024.791} {\bibfield  {journal} {\bibinfo
  {journal} {J. Fluid Mech.}\ }\textbf {\bibinfo {volume} {996}},\ \bibinfo
  {pages} {A13} (\bibinfo {year} {2024})}\BibitemShut {NoStop}%
\bibitem [{\citenamefont {Squire}(1933)}]{squire_stability_1933}%
  \BibitemOpen
  \bibfield  {author} {\bibinfo {author} {\bibfnamefont {H.~B.}\ \bibnamefont
  {Squire}},\ }\bibfield  {title} {\bibinfo {title} {On the stability for
  three-dimensional disturbances of viscous fluid flow between parallel
  walls},\ }\href {https://doi.org/10.1098/rspa.1933.0193} {\bibfield
  {journal} {\bibinfo  {journal} {Proceedings of the Royal Society of London.
  Series A, Containing Papers of a Mathematical and Physical Character}\
  }\textbf {\bibinfo {volume} {142}},\ \bibinfo {pages} {621} (\bibinfo {year}
  {1933})}\BibitemShut {NoStop}%
\bibitem [{\citenamefont {John Soundar~Jerome}\ and\ \citenamefont
  {Chomaz}(2014)}]{john_soundar_jerome_extended_2014}%
  \BibitemOpen
  \bibfield  {author} {\bibinfo {author} {\bibfnamefont {J.}~\bibnamefont {John
  Soundar~Jerome}}\ and\ \bibinfo {author} {\bibfnamefont {J.-M.}\ \bibnamefont
  {Chomaz}},\ }\bibfield  {title} {\bibinfo {title} {Extended {Squire}’s
  transformation and its consequences for transient growth in a confined shear
  flow},\ }\href {https://doi.org/10.1017/jfm.2014.83} {\bibfield  {journal}
  {\bibinfo  {journal} {J. Fluid Mech.}\ }\textbf {\bibinfo {volume} {744}},\
  \bibinfo {pages} {430} (\bibinfo {year} {2014})}\BibitemShut {NoStop}%
\bibitem [{\citenamefont {Oberlack}\ \emph {et~al.}(2026)\citenamefont
  {Oberlack}, \citenamefont {Wilhelm}, \citenamefont {Görtz}, \citenamefont
  {Conrad}, \citenamefont {Yalcin}, \citenamefont {De~Broeck},\ and\
  \citenamefont {Wang}}]{oberlack_new_2026}%
  \BibitemOpen
  \bibfield  {author} {\bibinfo {author} {\bibfnamefont {M.}~\bibnamefont
  {Oberlack}}, \bibinfo {author} {\bibfnamefont {K.~V.}\ \bibnamefont
  {Wilhelm}}, \bibinfo {author} {\bibfnamefont {S.}~\bibnamefont {Görtz}},
  \bibinfo {author} {\bibfnamefont {J.}~\bibnamefont {Conrad}}, \bibinfo
  {author} {\bibfnamefont {A.}~\bibnamefont {Yalcin}}, \bibinfo {author}
  {\bibfnamefont {L.}~\bibnamefont {De~Broeck}},\ and\ \bibinfo {author}
  {\bibfnamefont {Y.}~\bibnamefont {Wang}},\ }\bibfield  {title} {\bibinfo
  {title} {New subcritical oblique modes: {On} an extension of {Squire}'s
  theorem for spatiotemporally evolving modes},\ }\href
  {https://doi.org/10.1103/62zg-wqzy} {\bibfield  {journal} {\bibinfo
  {journal} {Phys. Rev. Fluids}\ }\textbf {\bibinfo {volume} {11}},\ \bibinfo
  {pages} {043902} (\bibinfo {year} {2026})}\BibitemShut {NoStop}%
\bibitem [{\citenamefont {Johnson}(1978)}]{johnson_numerical_1978}%
  \BibitemOpen
  \bibfield  {author} {\bibinfo {author} {\bibfnamefont {C.~R.}\ \bibnamefont
  {Johnson}},\ }\bibfield  {title} {\bibinfo {title} {Numerical {Determination}
  of the {Field} of {Values} of a {General} {Complex} {Matrix}},\ }\href
  {https://doi.org/10.1137/0715039} {\bibfield  {journal} {\bibinfo  {journal}
  {SIAM J. Numer. Anal.}\ }\textbf {\bibinfo {volume} {15}},\ \bibinfo {pages}
  {595} (\bibinfo {year} {1978})}\BibitemShut {NoStop}%
\bibitem [{\citenamefont {Kato}(1966)}]{kato_perturbation_1966}%
  \BibitemOpen
  \bibfield  {author} {\bibinfo {author} {\bibfnamefont {T.}~\bibnamefont
  {Kato}},\ }\href {https://doi.org/10.1007/978-3-662-12678-3} {\emph {\bibinfo
  {title} {Perturbation theory for linear operators}}}\ (\bibinfo  {publisher}
  {Springer Berlin Heidelberg},\ \bibinfo {address} {Berlin, Heidelberg},\
  \bibinfo {year} {1966})\BibitemShut {NoStop}%
\bibitem [{\citenamefont {Gustavsson}(1986)}]{gustavsson_excitation_1986}%
  \BibitemOpen
  \bibfield  {author} {\bibinfo {author} {\bibfnamefont {L.~H.}\ \bibnamefont
  {Gustavsson}},\ }\bibfield  {title} {\bibinfo {title} {Excitation of {Direct}
  {Resonances} in {Plane} {Poiseuille} {Flow}},\ }\href
  {https://doi.org/10.1002/sapm1986753227} {\bibfield  {journal} {\bibinfo
  {journal} {Stud Appl Math}\ }\textbf {\bibinfo {volume} {75}},\ \bibinfo
  {pages} {227} (\bibinfo {year} {1986})}\BibitemShut {NoStop}%
\bibitem [{\citenamefont {Shanthini}(1989)}]{shanthini_degeneracies_1989}%
  \BibitemOpen
  \bibfield  {author} {\bibinfo {author} {\bibfnamefont {R.}~\bibnamefont
  {Shanthini}},\ }\bibfield  {title} {\bibinfo {title} {Degeneracies of the
  temporal {Orr}-{Sommerfeld} eigenmodes in plane {Poiseuille} flow},\ }\href
  {https://doi.org/10.1017/S0022112089000819} {\bibfield  {journal} {\bibinfo
  {journal} {J. Fluid Mech.}\ }\textbf {\bibinfo {volume} {201}},\ \bibinfo
  {pages} {13} (\bibinfo {year} {1989})}\BibitemShut {NoStop}%
\bibitem [{\citenamefont {Reddy}\ and\ \citenamefont
  {Henningson}(1993)}]{reddy_energy_1993}%
  \BibitemOpen
  \bibfield  {author} {\bibinfo {author} {\bibfnamefont {S.~C.}\ \bibnamefont
  {Reddy}}\ and\ \bibinfo {author} {\bibfnamefont {D.~S.}\ \bibnamefont
  {Henningson}},\ }\bibfield  {title} {\bibinfo {title} {Energy growth in
  viscous channel flows},\ }\href {https://doi.org/10.1017/S0022112093003738}
  {\bibfield  {journal} {\bibinfo  {journal} {J. Fluid Mech.}\ }\textbf
  {\bibinfo {volume} {252}},\ \bibinfo {pages} {209} (\bibinfo {year}
  {1993})}\BibitemShut {NoStop}%
\bibitem [{\citenamefont {Boyd}(2013)}]{boyd_chebyshev_2013}%
  \BibitemOpen
  \bibfield  {author} {\bibinfo {author} {\bibfnamefont {J.~P.}\ \bibnamefont
  {Boyd}},\ }\href@noop {} {\emph {\bibinfo {title} {Chebyshev and {Fourier}
  spectral methods}}},\ \bibinfo {edition} {2nd}\ ed.\ (\bibinfo  {publisher}
  {Dover Publications},\ \bibinfo {year} {2013})\BibitemShut {NoStop}%
\bibitem [{\citenamefont {Huang}\ and\ \citenamefont
  {Sloan}(1994)}]{huang_pseudospectral_1994}%
  \BibitemOpen
  \bibfield  {author} {\bibinfo {author} {\bibfnamefont {W.}~\bibnamefont
  {Huang}}\ and\ \bibinfo {author} {\bibfnamefont {D.~M.}\ \bibnamefont
  {Sloan}},\ }\bibfield  {title} {\bibinfo {title} {The {Pseudospectral}
  {Method} for {Solving} {Differential} {Eigenvalue} {Problems}},\ }\href
  {https://doi.org/10.1006/jcph.1994.1073} {\bibfield  {journal} {\bibinfo
  {journal} {Journal of Computational Physics}\ }\textbf {\bibinfo {volume}
  {111}},\ \bibinfo {pages} {399} (\bibinfo {year} {1994})}\BibitemShut
  {NoStop}%
\bibitem [{\citenamefont {Weideman}\ and\ \citenamefont
  {Reddy}(2000)}]{weideman_matlab_2000}%
  \BibitemOpen
  \bibfield  {author} {\bibinfo {author} {\bibfnamefont {J.~A.}\ \bibnamefont
  {Weideman}}\ and\ \bibinfo {author} {\bibfnamefont {S.~C.}\ \bibnamefont
  {Reddy}},\ }\bibfield  {title} {\bibinfo {title} {A {MATLAB} differentiation
  matrix suite},\ }\href {https://doi.org/10.1145/365723.365727} {\bibfield
  {journal} {\bibinfo  {journal} {ACM Trans. Math. Softw.}\ }\textbf {\bibinfo
  {volume} {26}},\ \bibinfo {pages} {465} (\bibinfo {year} {2000})}\BibitemShut
  {NoStop}%
\bibitem [{Note2()}]{Note2}%
  \BibitemOpen
  \bibinfo {note} {Note that Ref. \protect \citep {weideman_matlab_2000} uses
  the representation $\protect \tilde {D}$ only for the 4th-order derivative in
  Eq. (\ref {eq:eigval equation numerics}), while we follow Ref. \protect
  \citep {huang_pseudospectral_1994} and use $\protect \tilde {D}$ for all the
  derivative terms on the left-hand side of the equation.}\BibitemShut {Stop}%
\end{thebibliography}%
\end{document}